\documentclass[pra,12pt,floatfix]{revtex4}
\usepackage{graphicx}
\usepackage{amsmath}

\begin{document}

\title{Fusion process of Lennard-Jones clusters:
global minima and magic numbers formation.}

\author{Ilia A. Solov'yov}
\altaffiliation[Permanent address:]{A. F. Ioffe Physical-Technical
Institute, Russian Academy of Sciences, Polytechnicheskaya 26,
St. Petersburg  194021, Russia}
\email[Email address: ]{ilia@th.physik.uni-frankfurt.de}
\author{Andrey V. Solov'yov}
\altaffiliation[Permanent address:]{A. F. Ioffe Physical-Technical
Institute, Russian Academy of Sciences, Polytechnicheskaya 26,
St. Petersburg  194021, Russia}
\email[Email address: ]{solovyov@th.physik.uni-frankfurt.de}
\author{Walter Greiner}
\affiliation{Institut f\"{u}r Theoretische Physik der Universit\"{a}t
Frankfurt am Main, Robert-Mayer str. 8-10, Frankfurt am Main 60054, Germany}

\begin{abstract}
We present a new theoretical framework for 
modelling the  fusion process of Lennard-Jones (LJ) clusters.
Starting from the initial tetrahedral cluster configuration,
adding new atoms to the system and absorbing its energy  
at each step, we find cluster growing paths up
to the cluster sizes of up to 150 atoms. We demonstrate that
in this way all known global minima structures
of the LJ-clusters can be found.
Our method provides an efficient tool for the calculation
and analysis of atomic cluster structure. With its use
we justify the magic number sequence for the clusters
of noble gas atoms and compare it with experimental
observations. 
We report the striking correspondence of the peaks in
the dependence on cluster size of the second derivative of
the binding energy per atom calculated for the
chain of LJ-clusters based on the icosahedral symmetry
with the peaks in the abundance mass spectra experimentally measured
for the clusters of noble gas atoms.
Our method serves an efficient alternative to the global optimization techniques
based on the Monte-Carlo simulations and it can be applied for the solution
of a broad variety of problems in which atomic cluster structure  is
important.
\end{abstract}


\maketitle
\section{Introduction}
\label{intro}

It is well known that the sequence
of cluster magic numbers carries essential
information about the electronic and ionic structure
of the cluster \cite{LesHouches}. Understanding of
the cluster magic numbers is often equivalent
or nearly equivalent to the understanding of
cluster electronic and ionic structure. A good example
of this kind is the observation of the magic numbers
in the  mass spectrum of sodium clusters
\cite{Knight84}. In this case, the magic numbers  
were explained by the delocalized electron shell closings 
(see \cite{MetCl99} and references therein).
Another example is 
the discovery of fullerenes, and, in particular, the $C_{60}$ molecule
\cite{Kroto85}, 
which was made by means of the carbon clusters mass spectroscopy.

The formation of a sequence of cluster magic
numbers should be closely connected to the mechanisms of
cluster fusion.
It is natural to expect that one can explain
the magic number sequence and find the most stable cluster isomers
by modeling mechanisms of  cluster assembly and  growth.
On the other hand, these mechanisms are of interest on their own, and
the correct sequence of the magic numbers found in such a simulation
can be considered as a proof of the validity of 
the cluster formation model.

The problem of magic numbers  is closely connected to the
problem of searching for global minima on the cluster multidimensional
potential energy surface. The number of local minima
on the potential energy surface increases exponentially with the growth
cluster size and is estimated to be of the order of $10^{43}$ for
$N=100$ \cite{LesHouches}. Thus, searching for global minima becomes
increasingly difficult problem for large clusters.
There are different algorithms and methods of the
global minimization, which have been employed for the global 
minimization of atomic cluster systems.

One of the most widely used global optimization methods calls the
simulated annealing \cite{Kirkpatrick,Wille,Ma93,Ma94,Tsoo,Laaarhoven}.
This method is an extension of
metropolis Monte Carlo techniques \cite{Kalos}.
In the simulated annealing one starts calculation from a
high energy state of the system and then cools the system down by
decreasing its kinetic energy. In standard application, the system is annealed using
molecular dynamics or Monte-Carlo based methods, but also more sophisticated
variants of this algorithm are used, such as gaussian density annealing and
analogues \cite{Ma93,Ma94,Tsoo}.

A related method is quantum tunnelling \cite{Finnila,Amara,Sylvain}.
This method attempts to reduce the effects of barriers
on the potential energy surface. This is done by
allowing the system to behave quantum mechanically,
leading to the possibility of tunnelling. However, the computational
expense of this method grows exponentially with increasing the number
of atoms in the system, what makes this method applicable for relatively
small clusters \cite{Maranas92,Maranas94}.

Another class of global optimization methods is based on smoothing
the potential energy surface \cite{Stillinger, Piela,Pillardy}.
Within the framework of this method,
a transformation is applied to the potential energy surface in order to decrease
the number of high lying local minima. The global minimum of the smoothed
potential energy surface, which is found by steepest descent energy
minimization method, or by Monte-Carlo search,
is then mapped back to the original surface.

Another successful method of global optimization is the basin-hopping
algorithm \cite{LesHouches,Wales,Leary}.
This method involves a potential energy transformation, which
does not change neither the global minimum, nor the relative energies of local minima
as it is usually done in the potential energy surface smoothing methods.
In this method, the potential energy surface in any point of the configuration
space is assigned to that of the local minimum obtained by the given
geometry optimization method and the potential energy surface is mapped onto
a collection of interpenetrating staircases with basins corresponding to the
set of configurations which lead to a given minimum after optimization.

Successful global optimization results for LJ clusters reported to date
were obtained with the use of genetic algorithms
\cite{Goldberg, Niesse,Gregurick,Deaven,Romero}. These methods
are based on the idea that the population of clusters evolves to low energy
by mutation and mating of structures with low potential energy.
There are different versions of genetic algorithm (see papers cited above).
One of them is seeding algorithm, or seed growth method \cite{Niesse,Gregurick}.
In this method it is assumed that the most stable cluster of $N$ particles
can be obtained from the most stable cluster of $N-1$ particle by a random
addition of one particle near the boundary of the cluster and then by optimizing
the structure with a given optimization method. Genetic algorithms and other
methods related to them are based on stochastic minimization technique,
and require no
quantum calculation or much information about the potential energy surface,
as the previously mentioned algorithms.

The new algorithm \cite{GrowProcPRL} which we describe in detail in
this work for the first time is based on the dynamic searching
for the most stable cluster isomers in the cluster fusion process.
We call this algorithm as the cluster fusion algorithm (CFA).
Our calculations demonstrate that the CFA can be considered
as an efficient alternative to the known techniques of the cluster 
global minimization. The big advantage of the CFA consists in 
the fact that it allows to study  not just the optimized cluster geometries,
but also their formation mechanisms.

In the present work we approach the formulated problem 
in a simple, but general, form. In our
simplest scenario, we assume that atoms in a cluster are bound 
by LJ potential
and the cluster fusion  takes place atom by atom.
At each step of the fusion process all atoms in the system
are allowed to move, while the energy of the system is
decreased. The motion of the atoms is stopped when the
energy minimum  is reached. The geometries and energies 
of all cluster isomers found in this way are stored
and analyzed.
On the first glance,
this method is somewhat similar to the genetic algorithms and to the seed
growth method, in particular.
However, the principle difference between the CFA and the genetic algorithms
consists in the fact that in CFA the new atoms are fused to the system not
randomly, but to certain
specific points in the system,
like the cluster centre of mass, the centres of mass of the cluster
faces or some specific points in the vicinity
of the cluster surface (see section \ref{model} for detail).
Such an approach, allows one
to model various physical situations and scenarios of the
cluster fusion process, which are beyond the scope of
the stochastic seed-growth method.
Thus, with the use of the CFA, we have determined 
the fusion paths for all known global energy minimum 
LJ cluster configurations in the size range up to $N \leq 150$.
We have also determined the fusion paths for the
chains of clusters based on the
icosahedral, octahedral, decahedral and tetrahedral symmetries.
We report the striking correspondence of the peaks in
the dependence on cluster size of the second derivative of
the binding energy per atom calculated for the
chain of LJ clusters based on the icosahedral symmetry
with the peaks in the abundance mass spectra experimentally measured
for the clusters of noble gas atoms \cite{Echt81,Haberland94,Harris84}.

Our paper is organized as follows.  In section \ref{model}, we 
describe theoretical model for the cluster fusion process
and the CFA.
In section \ref{results}, we present and discuss the results
of computer simulations based on the CFA.
In section \ref{conclusion}, we draw a conclusion to this paper.
In Appendix \ref{algorithms}, we provide additional details of
the numerical algorithms developed in our work.

\section{Theoretical model for cluster fusion process}
\label{model}

A group of atoms bound together 
by interatomic forces is called an atomic cluster (AC).
There is no qualitative distinction between small clusters
and molecules. However, as the number of atoms
in the system increases, ACs acquire more and more specific
properties making them unique physical objects different
from both single molecules and from the solid state.

Each stable cluster configuration corresponds to
a minimum on the multidimensional potential energy surface of
the system. The number of local minima
on the potential energy surface increases very rapidly 
with the growth cluster size. Thus, searching for 
the most stable cluster configurations possessing the absolute energy minimum,
the so-called global energy minimum structures,
becomes increasingly difficult problem for large clusters.

Searching for such global energy minimum cluster structures is one
of the focuses of our paper. This problem by itself is not new
as it is clear from introduction. However, in our work
we approach this problem at a new perspective and investigate
not just the global energy minimum cluster structure by its own, 
but rather the formation of such structures
in the cluster fusion process.

From the physical point of view, it is natural to expect 
that one can find the most stable cluster isomers by
modeling the mechanisms of cluster assembly and growth, i.e.
the cluster fusion process \cite{GrowProcPRL}. This idea is based on the fact
that in nature, or in laboratory, the most stable clusters
are often obtained namely in the fusion process.
This idea is general and is applicable to different types
of clusters. 

In this work we consider ACs formed by atoms 
interacting with each other via the pairing force.  
The interaction potential 
between two atoms in the cluster can, in principle, be arbitrary.
For concrete computations, we use the Lennard-Jones (LJ) potential,
\begin{equation}
U(r) = 4\varepsilon\left( \left(\frac{\sigma}{r}\right)^{12}
-\left(\frac{\sigma}{r}\right)^6\right),
\label{LJ_potential}
\end{equation}
\noindent
where $r$ is the interatomic distance, $\varepsilon$ is the
depth of the potential well ($\varepsilon > 0$), 
$2^{1/6} \sigma$ is the pair bonding length.
The constants in the potential allow one to model
various types of clusters for which LJ paring force approximation
is reasonable. The most natural systems of this kind are
the clusters consisting of noble gas atoms Ne, Ar, Kr, Xe
formed by van der Waals forces. 
The constants of the LJ potential appropriate for the noble gas
atoms one can find in \cite{RS}.
Thus, for Ne, Ar, Kr, Xe, $\varepsilon=3.6$, $12.3$, $17.2$, $24.3$ meV and
$\sigma =3.1$, $3.8$, $4.0$, $4.4$ $\AA$ respectively.
Note that for the LJ clusters it is always possible to choose 
the coordinate and energy scales so that $\sigma = 1$ and $\varepsilon = 1$. 
It makes all LJ cluster systems scalable. They only differ by
the choice of $\sigma$, $\varepsilon$ and
the mass of a single constituent (atom). In our paper we use LJ potential with
$\sigma=1$ and $\varepsilon=1/12$.

The AC treatment with the use of the LJ forces often implies 
the applicability of the classical Newton equations for the 
description of the AC dynamics. Following this line,
we describe the atomic motion in the cluster 
by the Newton  equations with the LJ pairing forces. 
In this case, the information about quantum properties of the system 
is hidden in the LJ-potential constants $\sigma$ and $\varepsilon$.
In computations, the system
of coupled Newton equations for all atoms in the cluster 
is solved numerically using the 4th order Runge-Kutta method.

\begin{figure}
\begin{center}
\includegraphics[scale=0.57]{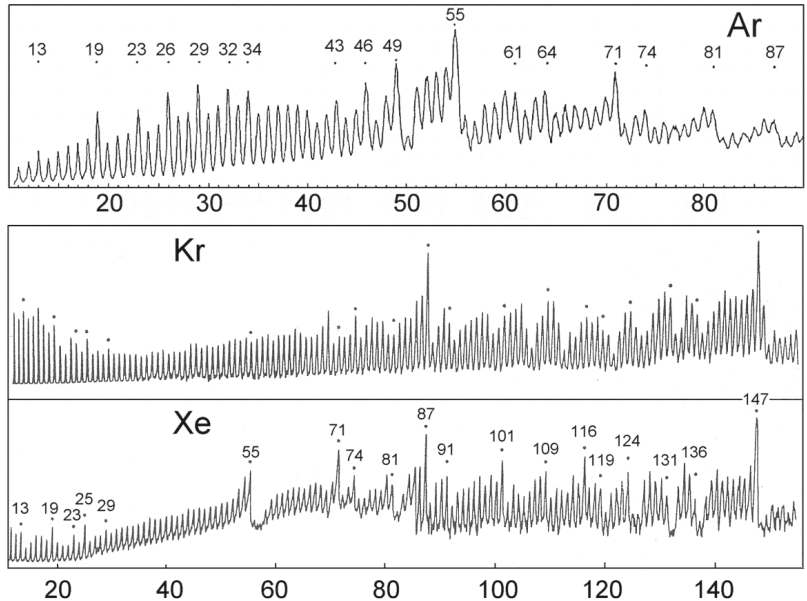}
\end{center}
\caption{
Experimentally measured abundance 
mass spectrum for the Ar, Kr and Xe
clusters \cite{Echt81,Haberland94,Harris84}.
}
\label{mass_spectra}
\end{figure}

Experimentally, mass spectra of the Ne, Ar, Kr, Xe clusters have been 
investigated in \cite{Echt81,Haberland94,Harris84}. 
In figure \ref{mass_spectra}, we compile the results of these 
measurements. This figure demonstrates that the mass spectra
for the clusters of noble gas atoms
have many common features. Some differences in the spectra
can be attributed to the differences in
the experimental conditions in which the spectra have been measured.

The peaks in the spectra indicate the enhanced
stability of certain clusters. These clusters are
called the magic clusters and their mass numbers $N$ are the magic numbers.
The enhanced stability of the magic clusters can have
different origin for various types of clusters. For LJ clusters, the 
origin of magic numbers is connected with the formation and filling
the icosahedral shells of atoms. The completed icosahedral shells 
of atoms correspond to the following sequence of magic numbers:
\begin{equation}
N = \frac{10}{3} z^3 +5 z^2 + \frac{11}{3}z + 1 
\label{ico_m_num}
\end{equation}
\noindent
as it is clear from a simple geometry analysis. 
Here, the integer $z= 1, 2, 3, 4 ...$ is
the order of the icosahedral shell. 
The first four icosahedral magic numbers, N, as they follow from
(\ref{ico_m_num}) are equal to 13, 55, 147, 309.
In figure \ref{mass_spectra}, one can see, however, many more peaks
corresponding to the formation of magic clusters having
the partially completed icosahedral shells.
In this paper, we elucidate the origin of all the peaks in the
recorded mass spectra of the noble gas ACs 
in the size range up to $N \leq 150$ by 
modeling the mechanisms of cluster fusion and by
finding in this way the most stable cluster isomers, i.e.
the global energy minimum cluster structures.

To solve this problem, for each $N$, we need to find solutions
of the Newton equations leading to the stable cluster configurations
and then to choose the one, which is energetically the most favourable.
The choice of the initial conditions for the simulation and the algorithm 
for the solution of this problem are described below.

Each stable cluster configuration corresponds to a
minimum on the cluster potential energy surface, i.e. to the point
in which all the atoms in the system are located in their equilibrium
positions. A minimum can be found by allowing atoms 
to move, starting from a certain initial cluster
configuration, and by absorbing all their kinetic energy
in the most efficient way.
If the starting cluster configuration for $N+1$ atoms
has been chosen on the basis of the global minimum
structure for $N$ atoms, then it is natural to assume, and 
we prove this in the present work, that very often the global minimum
structure for $N+1$ atoms can be easily found. The success
of this procedure reflects the fact that in nature 
clusters in their global minima often emerge, namely, in the
cluster fusion process, which we simulate in such calculation.

We have employed the following algorithm for the kinetic energy
absorption. At each step of the calculation
we consider  the motion 
of one atom only, which undergoes the action of the maximum
force.  At the point, in which the kinetic energy
of the selected atom is maximum, we set the absolute
value of its velocity to zero. This point corresponds
to the minimum of the potential well at which the selected atom moves.
When the selected atom is brought to the equilibrium position, 
the next atom is selected to move and  the
procedure of the kinetic energy absorption repeats.
The calculation stops when all the atoms are in equilibrium.

We have considered a number of scenarios of the cluster
fusion  on the basis of the developed algorithm 
for finding the stable cluster configurations.  

In the simplest scenario clusters of $N+1$ atoms are generated
from the $N$-atomic clusters by adding one atom 
to the system.
In this case the initial conditions for the
simulation of  $(N+1)$-atomic clusters 
are obtained on the basis of the chosen $N$-atomic cluster configuration by
calculating the coordinates of an extra atom fussed to the system
on a certain rule. We have probed the following paths:
the new atom can be fussed either

({\bf {\it A1}}) to the center of mass of the cluster, or 

({\bf {\it A2}}) randomly outside the cluster, 
but near its surface, or

({\bf {\it A3}}) to the centres of mass of 
all faces of the cluster, or

({\it A4}) to the points that are close 
to the centres of all faces of the cluster, located 
from both sides of the face on the perpendicular to it, or

({\it A5}) to the centres of mass of the faces laying 
on the cluster surface. 

Here, the cluster surface is considered  as a polyhedron, 
so that the vertices of the polyhedron are 
the atoms and two vertices are connected with an edge, 
if the distance between them is less than the value $d=1.2d_0$, where
$d_0 = 2 ^{1/6}\sigma$ is the bonding length of a free pair of atoms.
The cluster surface is considered as 
a polyhedron that covers all the atoms in the system
and has the minimum volume. The whole cluster volume is
divided on a sum of attached triangle, square and pentagonal
pyramids. In the {\it A3} method, all the faces in the
cluster are counted as a sum of the faces laying on the
cluster surface and the faces of the pyramids filling the cluster volume.

The choice of the method how to fuse atoms to the system depends
on the problem to be solved. 
The {\bf{\it A1}} method is appropriate in situations
when the new atoms are fused into the cluster volume.
The {\bf{\it A2}} simulates the process when atoms are fused
to the cluster surface.
By both  {\bf{\it A1}}  and {\bf{\it A2}} methods, 
large clusters consisting of 
many particles can be generated rather quickly. 
The {\bf{\it A2}} method 
is especially fast, because fusion of one atom to the boundary 
of the cluster usually does not lead to the recalculation of its central 
part. The {\bf{\it A3}} and {\bf{\it A4}} methods can be used
for searching the most stable, i.e. energetically favourable, 
cluster configurations or for finding  cluster isomers with some
other specific properties.
The {\bf{\it A4}} method leads to finding more cluster isomers than 
the {\bf{\it A3}} one, but it takes more CPU time. 
The {\bf{\it A5}} method is especially convenient for modeling the
cluster fusion process  which we focus on in this paper. Using this
method one can generate the cluster growth paths for the
most stable cluster isomers.

In the cluster fusion process, new atoms should be
added to the system starting 
from the initially chosen cluster configuration
step by step until the desired cluster size is reached. 
Each new step in the
cluster fusion process should be made with the use
of the methods {\bf {\it A1-A5}}. 
By methods {\bf{\it A3-A5}},
one generates many different cluster isomers
at each new step of the fusion process, those number 
grows rapidly as N increases. It is not feasible and often it is
not necessary to fuse new atoms to the all found isomers.
Instead, one can fuse atoms only to the selected clusters, which
are of interest.
Below, we outline a number of selection criteria, which can
be used for the cluster selection in the fusion process
depending on the problem to be solved.
At each new step of the fusion process:

({\bf{\it SE1}})  one of the clusters with 
the minimum number of atoms is selected, or 

({\bf{\it SE2}}) the cluster with the minimum energy 
among the already found stable clusters of the maximum size is selected, or
  
({\bf{\it SE3}}) a few low energy cluster isomers  
among the already found stable clusters of the maximum size are selected, or

({\bf{\it SE4}}) the cluster with the maximum energy 
among the already found stable clusters of the maximum size is selected, or

({\bf{\it SE5}}) the cluster possessing a significant structural
change among the already found stable clusters is selected.

The {\bf{\it SE1}} criterion 
is relevant in the situation, when the full search of
cluster isomers is needed. It is
applicable to the systems with relatively
small number of particles.
Note that each cluster can be selected only once.
The {\bf{\it SE2}} criterion is the most relevant for modelling
the cluster fusion process and the purposes of the present work. 
It turns out to be very efficient and often leads
to finding the most stable cluster configurations
for a given number of particles. 
However, this is not always the case, particularly for large clusters.
In these situations, the better results on the cluster global 
optimization can be achieved with the use of {\bf{\it SE3}}
method. However, this method takes more CPU time.
The {\it SE4} and {\it SE5} criteria
turn out to be useful for the redirection of the
cluster fusion process towards the lower energy
cluster isomer branches.

Calculations performed with the use of the methods
described above show that often clusters of higher
symmetry group possess relatively low energy. Thus, the symmetric
cluster configurations are often of particular interest.
The process of searching for the symmetric cluster configurations
can be speed up significantly, by the impose of
the symmetry constraints in the cluster fusion process.
This means that for obtaining a symmetric 
$N$ atomic cluster isomer
from the initially chosen symmetric $(N-M)$-atomic configuration one
should fuse $M$ atoms to the surface of this isomer symmetrically. 

This goal can be achieved by various methods.

({\bf{\it SY1}}) The planes of symmetry of the parent cluster are to be found.
Their intersections determine the axes of symmetry. 
The atoms of the parent cluster do not move and held their position, 
then an atom is added outside the cluster surface. 
The initial coordinates of it are not important, because its stable position  
in the vicinity of the frozen cluster surface is calculated. 
Then,  this atom is reflected many times with respect to the found symmetry 
planes, and the atoms are added on the places of its images. The 
obtained configuration is symmetric also. After that, 
all the atoms in the system are allowed to move while their 
kinetic energy is absorbed.
If the obtained stable cluster configuration 
is symmetric, it can be used 
as the initial cluster for further computations of this type. 
Using this procedure several times the chains of clusters of
certain symmetry can be found.

({\bf{\it SY2}}) This method is similar to the one described above,
but now the atoms are added to the centres of the faces on the cluster
surface. If the cluster 
has two (or more) different types of faces, then, at first, 
the atoms are added 
to the centres of the faces of the first type and their
positions are optimized with frozen cluster core.
Then, the whole system is optimized and checked for symmetry. 
If it is symmetric, the process continues, 
atoms are added to the faces of the second type 
and the same actions are performed. 
This method makes possible to find more stable configurations than the 
first one.

({\bf{\it SY3}}) The atoms are added to the axes of symmetry outside 
the cluster surface.
Then, the atoms of the parent cluster are assumed to be frozen 
and the optimized coordinates of the added atoms are
calculated. After that the optimization 
of the whole system is performed.  
If the obtained cluster configuration is symmetric, 
it can be used as the initial cluster for further computations.

Note that the {\bf{\it SY1}}-{\bf{\it SY3}} methods 
do not model any concrete physical scenario for the cluster fusion process.
They rather represent efficient mathematical algorithms for the generation
symmetrical cluster configurations, which often turn out to be 
very useful for the cluster structure analysis.

Note also that in addition to the cluster fusion techniques described above
any stable cluster configuration can be manually modified by
adding or removing a number of atoms. These modifications
can also be performed using advantages of the cluster symmetry
if it does exist. The manual modification of the cluster configuration
is useful when the cluster geometry is already established
or nearly established and only small modifications of the system
are required. For example, by this method one can perform 
the surface rearrangement of atoms leading  to the growth
the number of bonds in the cluster and thus to
the enhancement of its stability. This goal can be achieved by
the replacement of a few loosely bound atoms on the cluster surface to
the positions, in which they have larger number of bonds.
As it is demonstrated in the
next section, such rearrangements result in finding
the cluster isomers, which alternatively can be obtained when
performing the fusion simulation with the use 
of the {\bf{\it SE3}} criterion.
The manual modifications technique requires usually
much less CPU time.
Cluster configurations found in  this way can be
used as starting configurations for the subsequent 
cluster fusion process.

\section{Results and discussions}
\label{results}

Using our algorithms, we have examined various paths 
of the cluster fusion process, 
and determined the most stable cluster isomers for the cluster
sizes of up to 150 atoms.

We have generated the chains of clusters 
based on the icosahedral, octahedral, tetrahedral and decahedral
symmetries with the use of the {\it A3-A5} and {\it SE2-SE5}
methods. We show that clusters possessing
icosahedral and decahedral type of lattice are energetically more
favourable.
For the calculated cluster chains, we  
analyze the average interatomic distances in clusters,
average number of bonds and binding energies.
Using these results we explain the sequence of magic numbers, experimentally
measured for the noble gas clusters, and elucidate the level of applicability
of the liquid drop model for the description of the
LJ cluster systems. Using the
{\it A1-A2} methods we calculate chains of energetically unfavourable
cluster isomers as an example of the spontaneous cluster
growth of such systems.
Also, we consider the formation of symmetrical
clusters using methods {\it SY1-SY3} and their correspondence
to the global energy minimum cluster structures.

\subsection{Fusion of global energy minimum clusters}
\label{fusion}
\subsubsection{LJ cluster geometries}

\begin{figure}
\begin{center}
\includegraphics[scale=0.74]{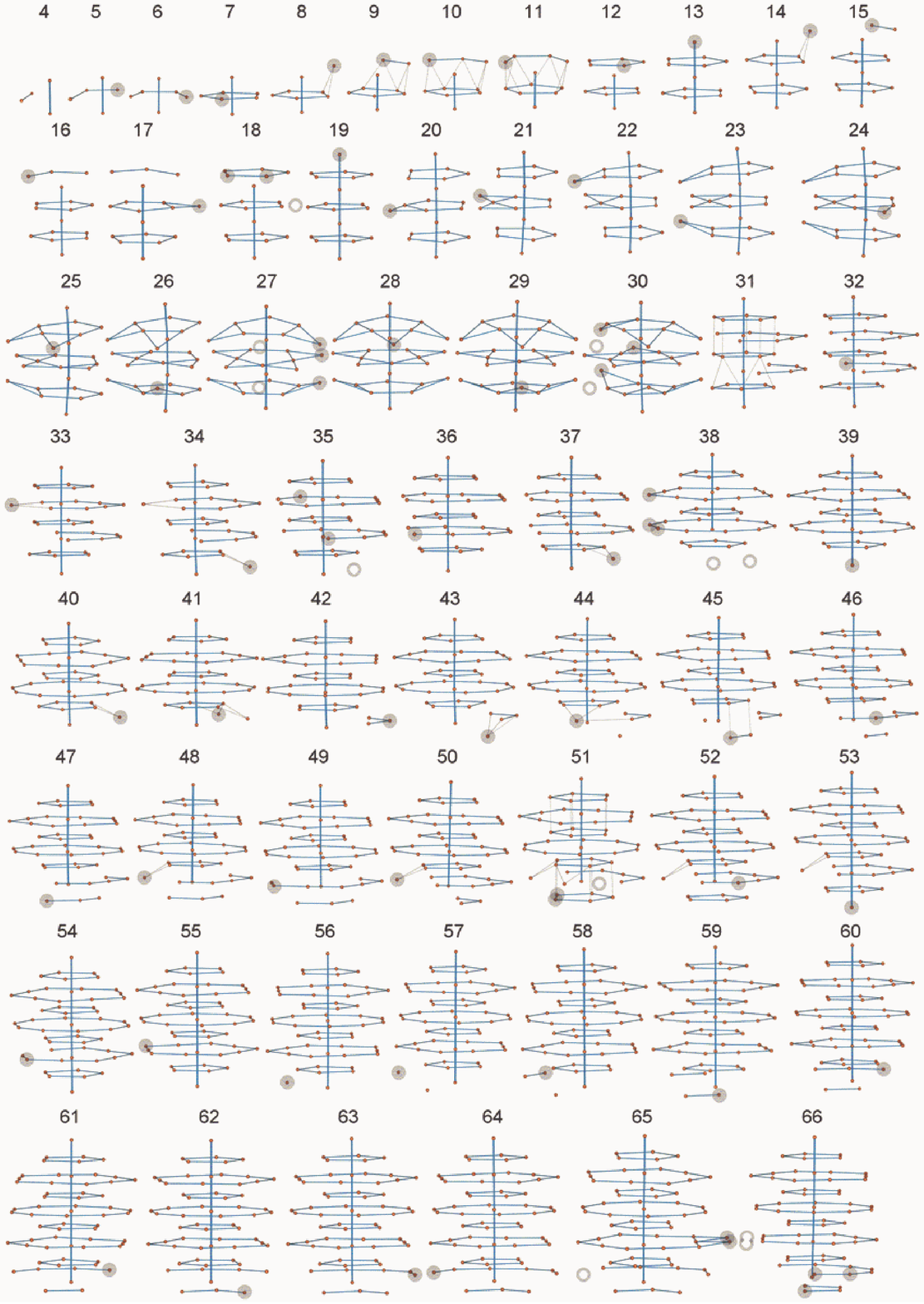}
\end{center}
\end{figure}

\begin{figure}
\begin{center}
\includegraphics[scale=0.74]{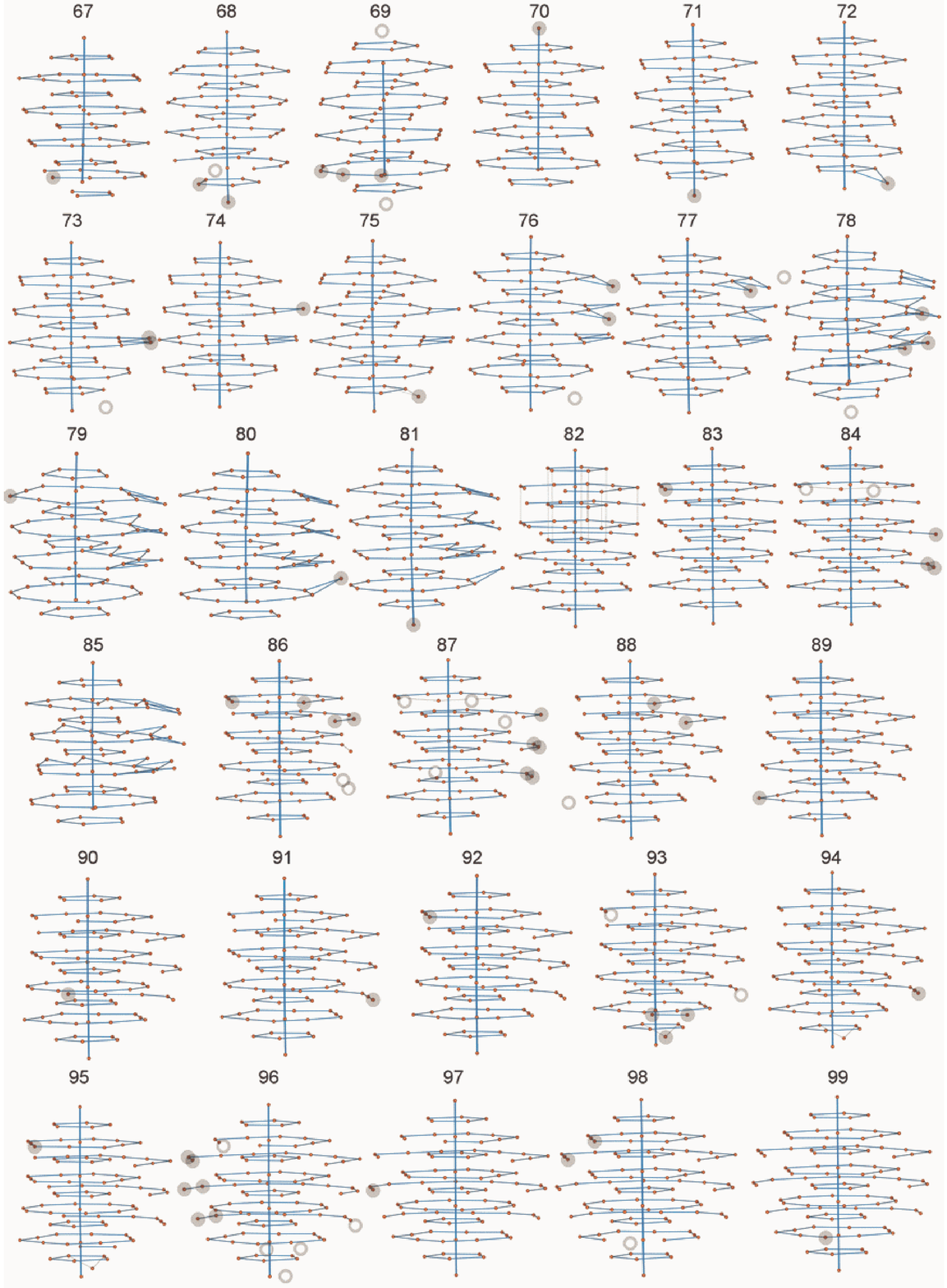}
\end{center}
\end{figure}

\begin{figure}
\begin{center}
\includegraphics[scale=0.74]{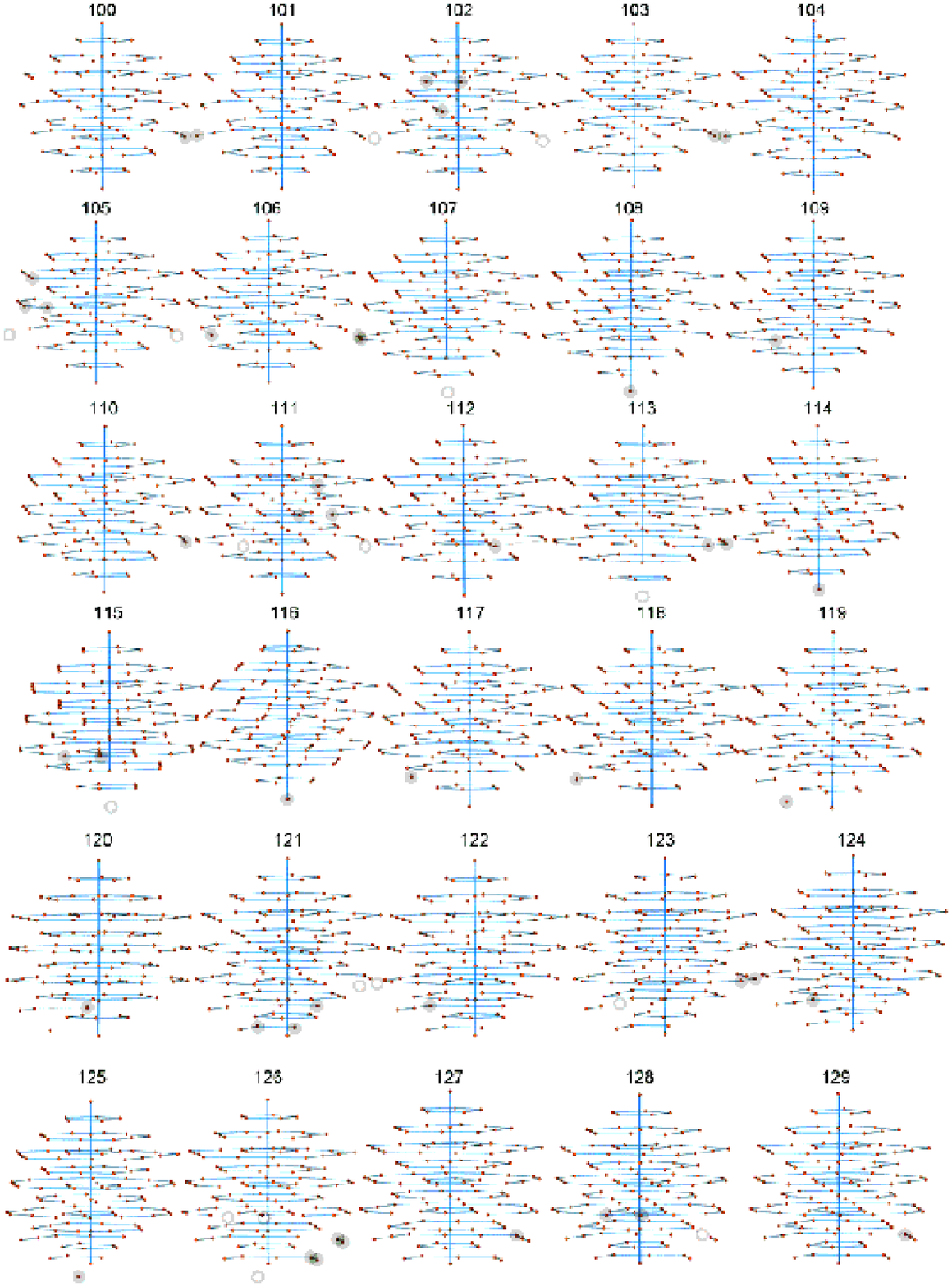}
\end{center}
\end{figure}

\begin{figure}
\begin{center}
\includegraphics[scale=0.74]{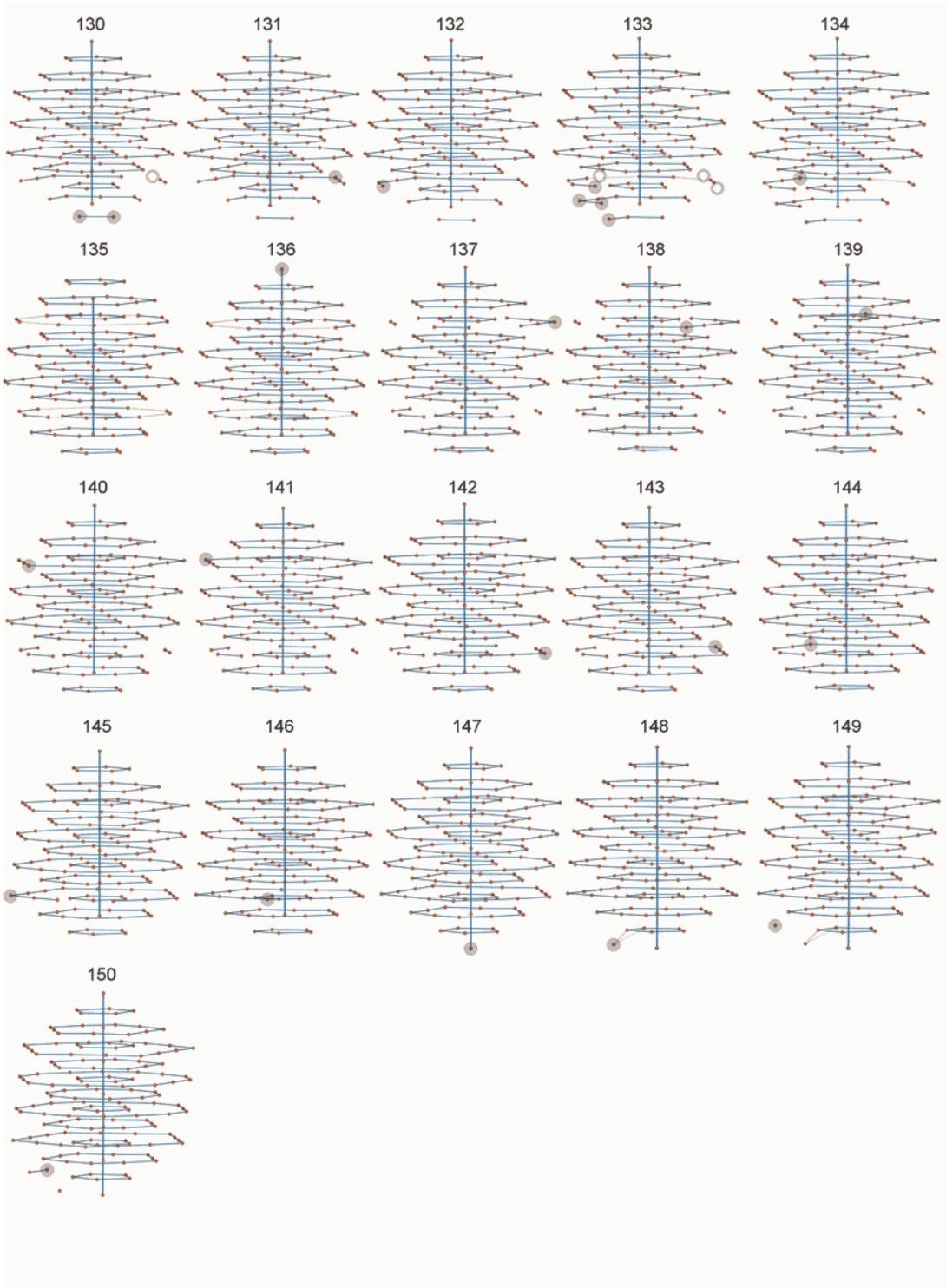}
\end{center}
\caption{
Growth of Lennard-Jones global energy minimum cluster structures based on the icosahedral
type of packing.
$LJ_4 - LJ_{66}$ (part a), $LJ_{67} - LJ_{99}$ (part b),
$LJ_{100} - LJ_{129}$ (part c) and $LJ_{130} - LJ_{150}$ (part d).
The new atoms added to the cluster are marked by grey circles, while
grey rings show the atoms removed.
}
\label{geometry}
\end{figure}

The growth of the most stable, i.e. possessing the lowest energy, LJ clusters
based on the icosahedral symmetry,
the so-called global energy minimum cluster structures, is
illustrated in figure \ref{geometry}, parts a-d. We analyse the cluster
geometries within the size range $4 \leq N \leq 150$ and determine the
transition path from smaller clusters to larger ones.
The cluster geometries have been determined using the
methods described in section \ref{model}.
Namely, we have used {\it A3-A5} methods combined with {\it SE2-SE5} cluster
selection criteria. 

For each cluster, we show the cluster axis, i.e.
a group of atoms located
on a line, that goes through the cluster centre of mass.
The cluster axis can be linear 
(see the clusters with higher symmetry, 
for example, $LJ_{13}$, $LJ_{55}$, $LJ_{71}$, $LJ_{147}$)
or slightly distorted (see, for example, non-completed
cluster configurations $LJ_{25}$, $LJ_{27}$,
$LJ_{76}$, $LJ_{116}$). Other atoms in the cluster are located around
the cluster axis and form the completed or open polygonal rings (see
figure \ref{geometry}).
The bonds between the atoms laying in the axis and in the rings
are shown by thick and thin lines respectively.
In some particular cases, we present additional connections in order
to stress specific structural
properties of some clusters (see, for example, $LJ_{11}$, $LJ_{31}$,
$LJ_{51}$, $LJ_{82}$).

Figure \ref{geometry} demonstrates how the cluster of a
certain size can be obtained from the smaller one.
For each cluster, we draw the added atoms by grey circles.
For some clusters, it is necessary additionally to replace one atom,
or even a few atoms from one position to another in order to reach the
global energy  minimum.
In these cases, we mark the positions, from which the atoms are removed,
by grey rings
(see, for example, $LJ_{96}$, $LJ_{102}$, $LJ_{126}$).

Figure \ref{geometry} demonstrates that the majority of
energetically favourable cluster structures
can be obtained with the use of the {\it A5} method combined
with the {\it SE2} cluster selection criterion, according to which
the global energy
minimum cluster geometry is obtained from the preceding cluster
configuration by fusion a single atom to the cluster surface.
Such situation takes place for 
$N=$ 4-17, 19-26, 28-29, 32-34, 36-37, 39-50, 52-64, 67, 70-72,
74-75, 77, 79-81, 83, 89-92, 94-95, 97, 99-101, 103-104, 106-110,
112, 114, 116-120, 122, 124-125, 127, 129, 131-132, 134, 136-150, i.e.
in about $75 \%$ cases.

The global energy minimum cluster structures presented in figure \ref{geometry}
coincide with those found by other cluster global optimization techniques
(see \cite{LesHouches,Wales,Leary} and references therein).
Note that the different choice of the constants $\varepsilon$ and $\sigma$
in various calculation schemes
does not change the cluster geometry and influences only the scale of the
system's size and its total energy.
However, the existing global optimization techniques
do not allow to perform the cluster growth
analysis, which we carry out in our work for the first time.
Some preliminary results of our work
have been published in \cite{GrowProcPRL, NanoMeeting}.

\begin{figure}
\begin{center}
\includegraphics[scale=0.50]{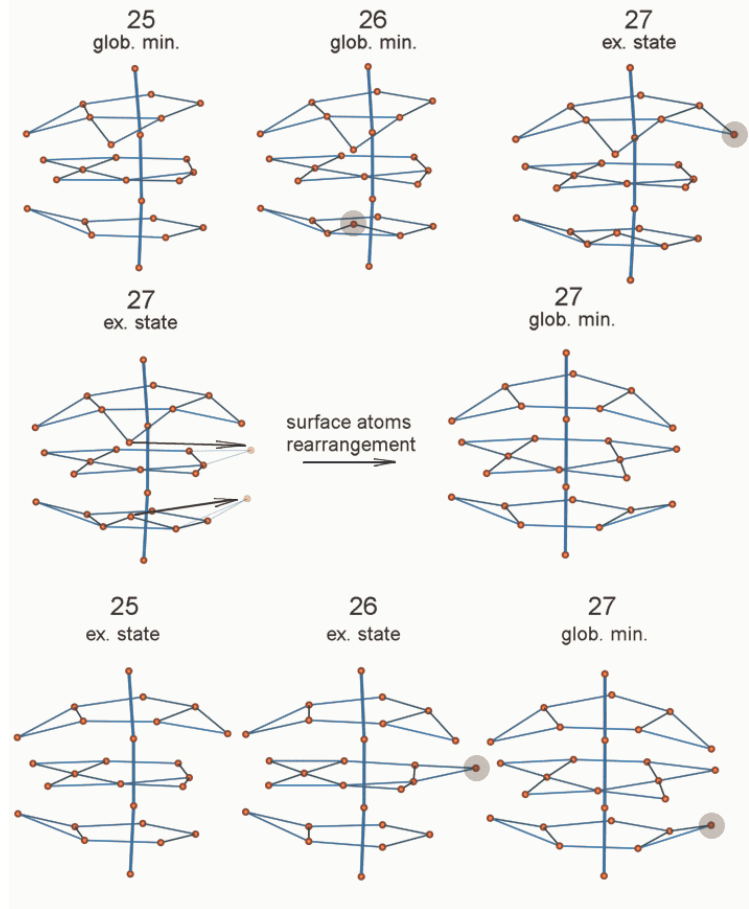}
\end{center}
\caption{
Fusion of a single atom to the global energy minimum cluster structure
of 26 atoms does not lead to the global energy minimum of the $LJ_{27}$
cluster (first row). Rearrangement of surface atoms in the $LJ_{27}$ cluster
leading to the formation of the global energy minimum cluster structure is
needed (second row). The result of such rearrangement can be obtained if one
starts the growth from the excited state
cluster isomer of the $LJ_{25}$ cluster (third row).
}
\label{rearrangement}
\end{figure}

It is interesting that all the calculated cluster geometries have the
structure, in which a number of completed and open polygons round the
cluster axis.
The maximum number of atoms in polygons depends on the cluster size.
Within the size range of $N\leq 150$, as it is seen from figure
\ref{geometry}, the pentagonal, decagonal and
pentadecagonal polygons are possible. The
fact that this type of rings within the cluster structure dominates
is closely connected with the prevalence of the icosahedral type
of packing of atoms in the LJ clusters.

Note that only the clusters possessing relatively high symmetry have all the
polygons rounding the axis completed.
For the most of the clusters, the polygons laying
close to the cluster surface are open
(see e.g. $LJ_{61}$, $LJ_{89}$, $LJ_{124}$). For such clusters, the atoms
added to the system occupy the positions in the open polygons
located near the cluster surface.

\subsubsection{Surface atoms rearrangements}

Our simulations demonstrate that the fusion of a single atom to the global
energy minimum cluster structure of $N$ atoms, in some cases
does not immediately lead to a global energy minimum of $(N+1)$-atomic cluster.
This happens for $N=$ 18, 27, 30, 35, 38, 51, 65,
66, 68, 69, 73, 76, 78, 84, 86, 87, 88, 93, 96, 98, 102, 105, 111, 113, 115,
121, 123, 126, 128, 130, 133, 135.
The global energy minimum cluster structure for these cluster sizes can not be found
directly with the use of {\it A5} method combined with the
{\it SE2} criterion from the global energy minimum cluster configuration of the smaller size.
In these cases the {\it SE2} criterion leads to finding the higher energy cluster
isomers, which we also call the excited state cluster isomers.
In order to determine the global energy minimum cluster configurations for the above
mentioned $N$, it is necessary to rearrange additionally one
or a few atoms in the excited state cluster isomer (see example shown in
figure \ref{rearrangement}).
In figure \ref{geometry}, we illustrate this procedure by marking the positions
of the rearranged atoms by grey circles. The initial positions of the atoms
are marked by grey rings.
This figure shows that the rearrangement takes place always
in the vicinity of the cluster surface.
This fact has a simple physical explanation.
The surface atoms are bound weaker than those inside the cluster volume,
and thus they are movable easier and allow the favourable
surface atomic rearrangement. This consideration demonstrates that
the surface rearrangement of atoms is an essential component of the
cluster growth process.

We illustrate the surface rearrangement of atoms in figure \ref{rearrangement}.
The first row in figure \ref{rearrangement} shows that
the fusion of a single atom to the global energy minimum cluster structure
of 26 atoms does not lead to the global energy minimum of the $LJ_{27}$
cluster. Thus, the rearrangement of surface atoms in the $LJ_{27}$ cluster
leading to the formation of the global energy minimum cluster structure is
needed. The necessary rearrangement of atoms is shown in the second row of
figure \ref{rearrangement}. The result of such rearrangement can be obtained
if one starts the cluster growth from the excited state
cluster isomer of the $LJ_{25}$
cluster, as it is illustrated in the third row of figure \ref{rearrangement}.

In this particular example, only the two smaller cluster sizes are involved in
the fusion process of the $LJ_{27}$ cluster. However, the cluster
fusion via excited states is not always that simple and evident.
In some cases, it involves more than 10 intermediate steps. Such situation
takes place, for example, for the
$LJ_{69}$ cluster, which can be obtained from the excited state isomer of
the $LJ_{55}$ cluster.

\subsubsection{Average number of bonds in LJ clusters}

\begin{figure}
\begin{center}
\includegraphics[scale=0.63,angle=270]{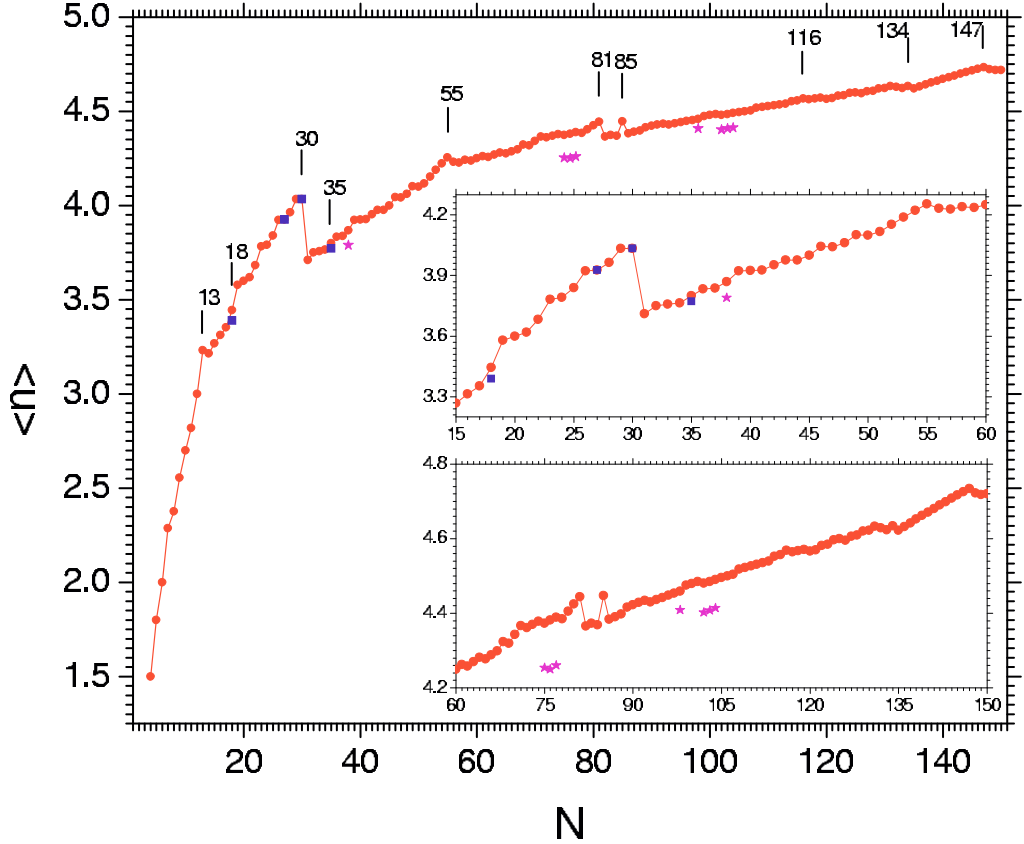}
\end{center}
\caption{
Average number of bonds in the icosahedral LJ global energy minimum cluster
structures as a function of cluster size (circles). Squares show that for the
close energy excited state cluster isomers and stars for the non-icosahedral
global energy minimum cluster structures. 
In the insets we show in greater scale
the average number of bonds in the regions $15\leq N\leq 60$ and
$60\leq N\leq 150$.
}
\label{bond_num}
\end{figure}

The rearrangement of the surface atoms can be performed algorithmically
by freezing almost all the atoms and allowing only a few atoms
on the cluster surface to move. However, the number of possible cluster
configurations increases rapidly with the growth cluster size,
and it becomes not feasible to optimize and investigate all of them.
Thus, one has to use a simple criterion for the selection of
some clusters, for which the surface rearrangement of atoms is of
interest. For such an analysis, we have introduced the criterion
based on counting the number of bonds in the cluster.
We consider the two atoms as bound if the
distance between them is smaller than the given cut-off distance value $d$.
Defining bonding
between the atoms in such a way, we then calculate the total
number of bonds in the cluster. This characteristic
can be used in searching for the required cluster
rearrangement, because the cluster with the maximum number of bonds,
possesses the highest binding energy.

In figure \ref{bond_num}, we present the dependence of average number of bonds
in the cluster as a function of the cluster size. For this calculation we have
used
the cut-off value $d=1.2d_0$, where $d_0=2^{1/6}\sigma$ is the
LJ potential bonding length.
In figure \ref{bond_num}, circles and squares show the average number
of bonds for the global energy minimum structures presented
in figure \ref{geometry} and for the excited state cluster isomers
with the energy close to the global minimum correspondingly.
This figure demonstrates that
circles always show the upper limit for the average number
of bonds in the cluster.

For some clusters, like $LJ_{27}$ or $LJ_{30}$, the number of
bonds in the global energy minimum cluster structure and the
closest excited state cluster isomer appears to be very close.
Such situations occur, when the energies of
the two clusters turn out to be very close.
In such cases, bonding between atoms at distances larger
than the chosen cut-off value becomes important and needs to
be counted in order to see that the number of bonds in the global energy
minimum cluster structure is maximal.
For example, the total number of bonds for the
two different cluster isomers of $LJ_{30}$ calculated
with the cut-off value $d$ is equal to $121$. 
Doubling the value of $d$ results in the total number of bonds
$343$ and $339$ for the global energy minimum and the excited state
cluster isomer respectively.

Such an analysis turns out to be very useful for the algorithmic search
of the surface rearrangement of atoms leading to the growth of the
number of bonds in the cluster and its global optimization.
In simple cases, the favourable
rearrangement is often obvious and it can be performed manually, by
the replacement of
a group of atoms from one position to another.
This is often the case for larger clusters,
see, for example, the rearrangement in the $LJ_{38}$ or $LJ_{69}$ clusters.

Note that the average number of bonds in the clusters with icosahedral
symmetry converges to the bulk limit relatively slowly. As it is
clear from a simple geometrical analysis, for an infinitively large
icosahedron the average number of bonds should be equal to $6$,
while for $LJ_{55}$, $LJ_{147}$ and $LJ_{309}$ it is 4.25, 4.73
and 5.00 respectively.

\subsubsection{LJ cluster lattices}

The methods and criteria described above appear to be insufficient
for the complete description of the LJ cluster fusion process.
At certain cluster sizes, a radical rearrangement of the cluster structure is necessary.
Below, we call such rearrangements as the core or lattice
rearrangements. Indeed, clusters in the fusion chain form the lattice
of a certain type. Different lattices are based on
different principles of the cluster packing and symmetry.
For LJ clusters, we analyzed
four different lattices: icosahedral, decahedral, octahedral and tetrahedral.
The cluster configurations found in our simulations
have been distinguished between the four groups according to their core symmetry.
The cluster core is formed by a certain number of completed shells of atoms.
Clusters consisting of the completed shells only are called the principal magic
clusters. The mass numbers for the principal magic clusters
(principal magic numbers) with the
icosahedral, decahedral, octahedral and tetrahedral type of packing
can be easily determined from simple geometric consideration and read as:

\begin{figure}
\begin{center}
\includegraphics[scale=0.72]{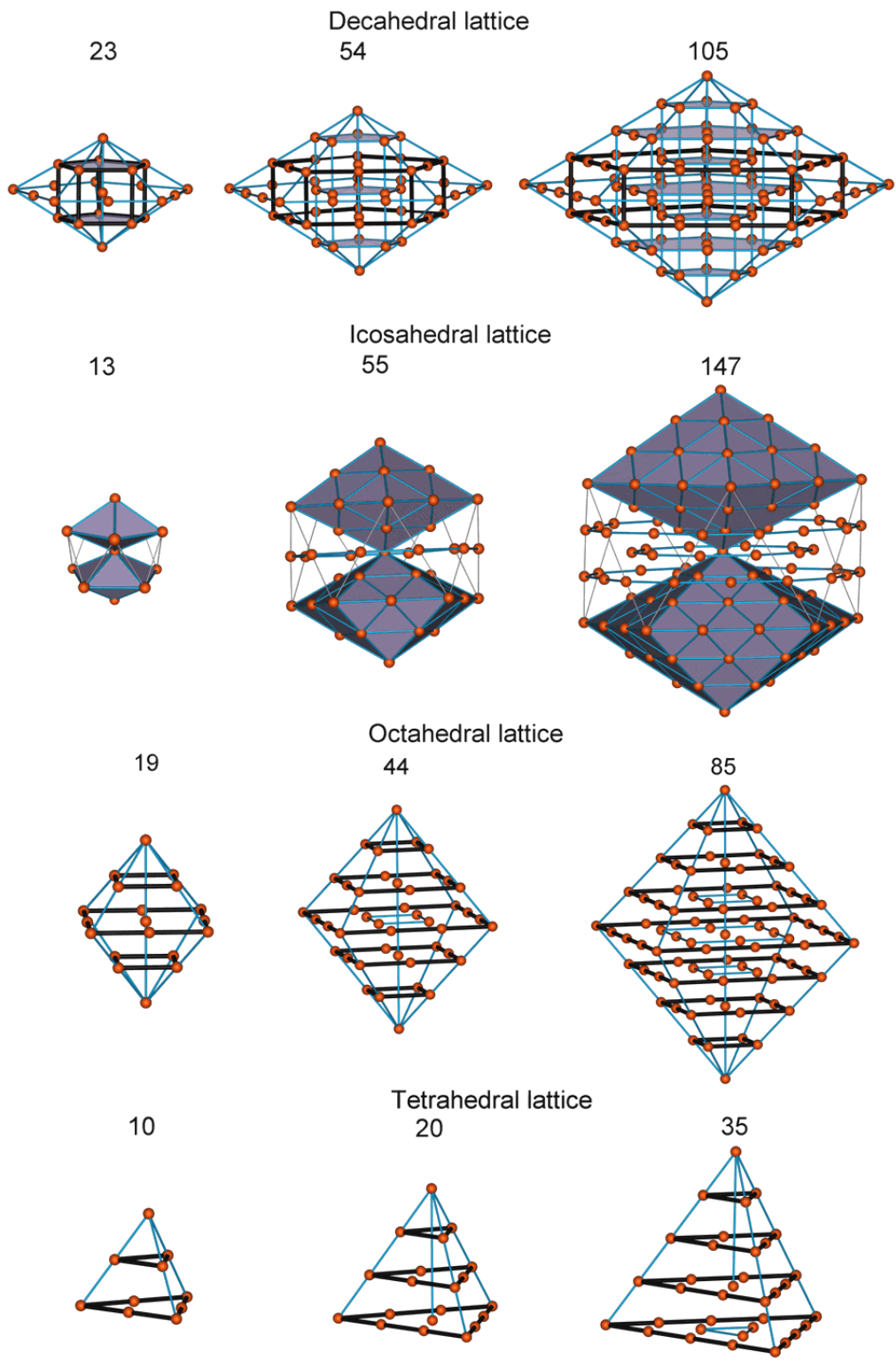}
\end{center}
\caption{
Geometries of magic clusters with decahedral lattice (first row),
icosahedral lattice (second row), octahedral lattice (third row),
tetrahedral lattice (fourth row).
}
\label{lattices}
\end{figure}

\begin{eqnarray}
N_{i} &=& \frac{10}{3} z^3 +5 z^2 + \frac{11}{3}z + 1\\
\nonumber
N_{d} &=& \frac{5}{6} z^3 +\frac{5}{2} z^2 + \frac{8}{3}z + 1\\
\nonumber
N_{o} &=& \frac{2}{3} z^3 +2 z^2 + \frac{7}{3}z + 1\\
\nonumber
N_{t} &=& \frac{1}{6} z^3 +1 z^2 + \frac{11}{6}z + 1\\
\nonumber
\label{m_num}
\end{eqnarray}

\noindent
where $N_{i}$, $N_{d}$, $N_{o}$ and $N_{t}$ are the principal
magic numbers for the icosahedral, decahedral, octahedral
and tetrahedral cluster packing, $z$ is the number of shells in the
cluster or its order.

In figure \ref{lattices},
we show the geometrical structure of a few principal magic clusters of each type. 
All the clusters are formed by polygons rounding
the cluster axis. Their number and type varies for different clusters.

The decahedral principal magic clusters are formed by pentagonal rings, that
are located one above another along the cluster axis
(see row 1 in figure \ref{lattices}). The number of
rings and their size depends on the number of shells, $z$.
Note that in decahedral clusters, the lines connecting corresponding
vertices of pentagonal rings of the same size are parallel to the
cluster axis.

The icosahedral principal magic clusters differ from the decahedral ones
(see row 2 in figure \ref{lattices}). Depending on the order of cluster,
the icosahedral cluster can additionally to pentagonal rings contain
decagonal, pentadecagonal and etc ones.
These additional rings glue together two equal decahedrons of the
same order having the common axis, along which the decahedrons
are turned one with respect to another on the angle $36^o$.
As a result, some of neighbouring rings in the icosahedral clusters
are also turned one with respect to another (see figure \ref{lattices}). 

In row 3 and 4 of figure \ref{lattices}, we show the geometries
of the principal octahedral and tetrahedral magic clusters
respectively. In
clusters of this type, squares and triangles correspondingly round the cluster
axis. 

\subsubsection{LJ cluster lattice rearrangements}

Let us now consider the role played by the lattice rearrangement in the
formation of the global energy minimum cluster structures. Figure
\ref{geometry} demonstrates that the global energy minimum
for $LJ_{31}$, $LJ_{82}$ and $LJ_{85}$ can not be found by the surface
rearrangement of atoms.

\begin{figure}
\begin{center}
\includegraphics[scale=0.80]{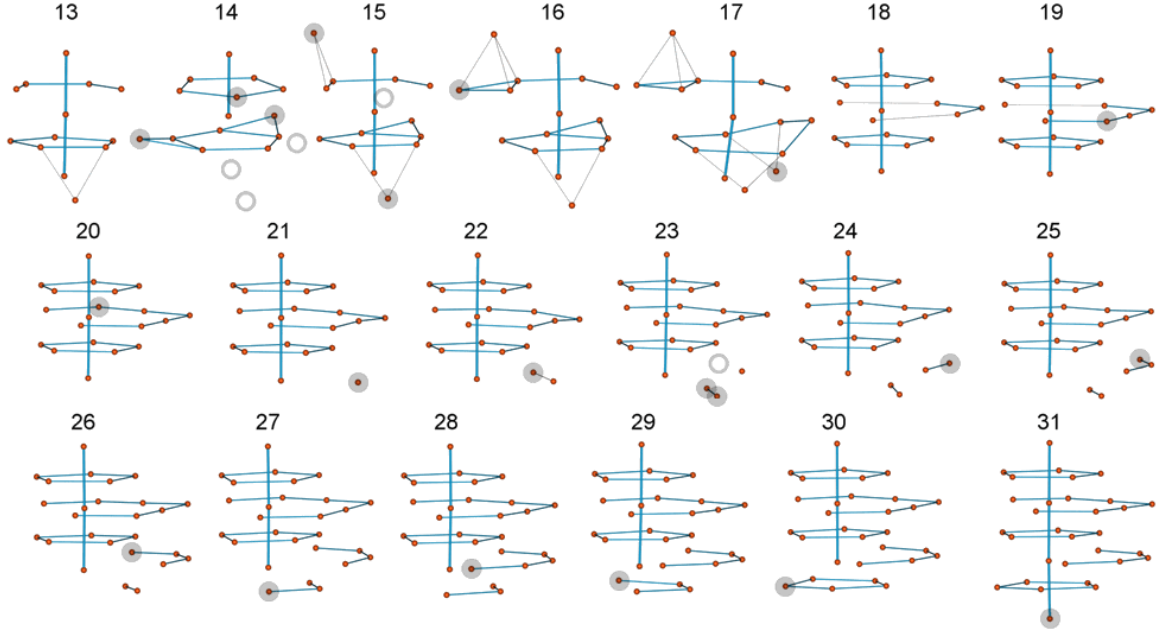}
\end{center}
\caption{
Formation of the $LJ_{31}$ global energy minimum structure takes place
via a chain of excited state cluster isomers with $N\geq 13$.
New atoms added to the system are marked by grey circles, while
grey rings demonstrate the atom removal.}
\label{13_to_31}
\end{figure}

Figure \ref{geometry}
clearly demonstrates that the $LJ_{31}$ cluster has obvious elements
of the decahedral packing. In order to stress this fact,
we connect the corresponding
vertices of the pentagons by thin lines. In the clusters of the smaller size,
the neighbouring pentagonal rings (completed or uncompleted)
are turned one with respect to another.
This feature is characteristic for the icosahedra type of cluster packing.
The $LJ_{31}$ cluster
structure arises in the cluster growing process via a long chain
of excited state cluster isomers with $N \geq 13$.
This chain is shown in figure \ref{13_to_31}. This figure demonstrates
that the essential
rearrangement occurs already in the $LJ_{18}$ cluster leading to the
formation of the
two co-ordinated pentagonal rings located one above another. After that
point, the formation of the $LJ_{31}$ cluster takes place
in a regular way and can be generated atom by atom.

\begin{figure}
\begin{center}
\includegraphics[scale=0.72]{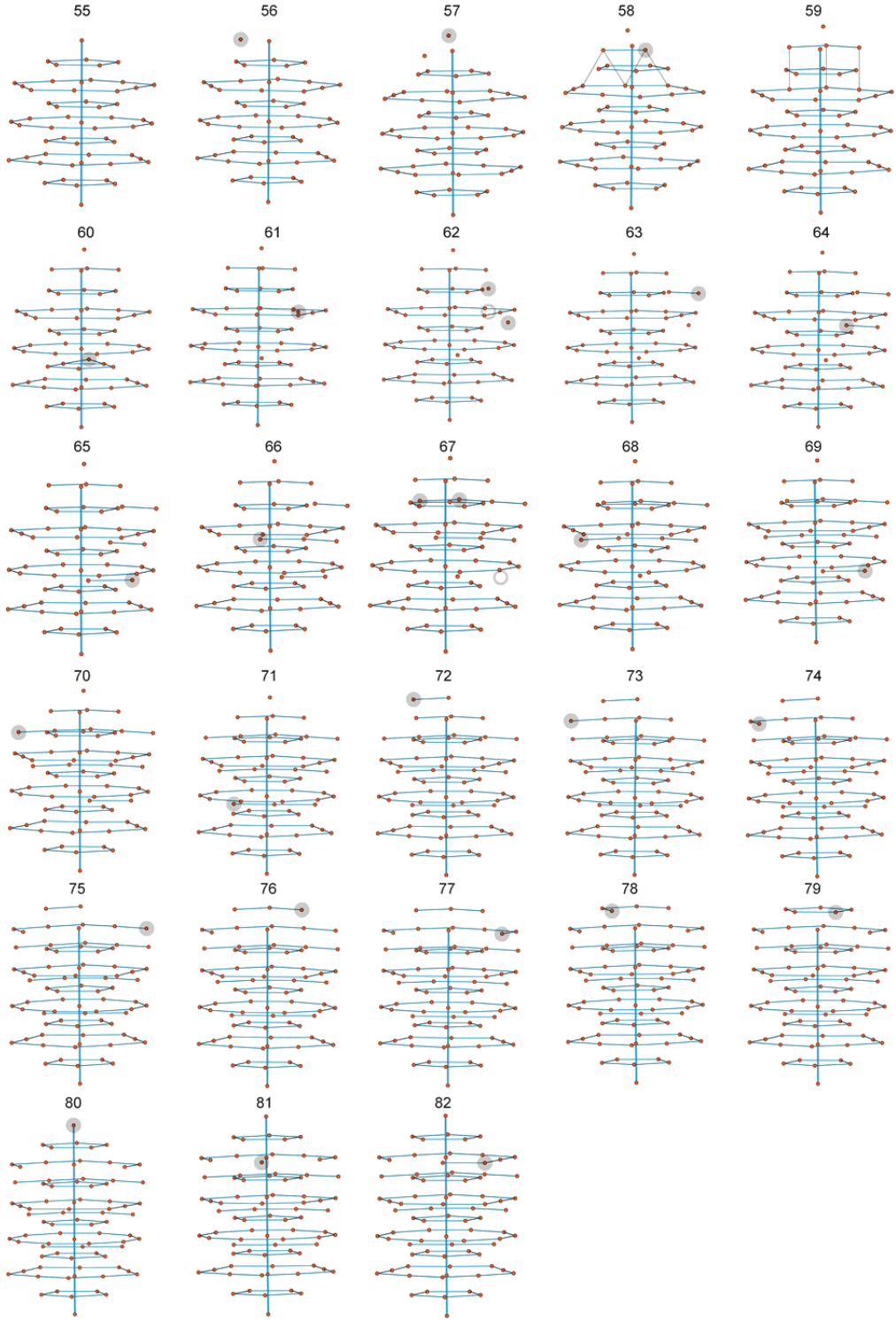}
\end{center}
\caption{
Formation of the $LJ_{82}$ global energy minimum structure involves
a chain of excited state cluster isomers with $N\geq 55$.
Atoms added to the system are marked by grey circles, while
grey rings demonstrate the atom removal.}
\label{55_to_82}
\end{figure}

The next prominent rearrangement of the cluster core takes place for
the $LJ_{82}$ cluster. The formation of this cluster involves
a long chain of excited state cluster isomers
with $N \geq 55$. In figure \ref{55_to_82},
we present this chain of clusters. The core structure
for the $LJ_{82}$ cluster
emerges in the $LJ_{59}$ cluster, in which the three atoms from
one pentagonal ring are located just above the
three atoms from another pentagonal
ring (see figure \ref{55_to_82}). The further formation of the $LJ_{82}$
cluster happens in a regular manner atom by atom.

The next cluster that can not be found from the previous one by the
surface rearrangement of atoms is the $LJ_{85}$ cluster. However, this
cluster configuration can be easily generated
starting from $LJ_{81}$ (see
figure \ref{geometry}, part b) and using a simple rearrangement
of atoms located at the cluster surface.

\begin{figure}
\begin{center}
\includegraphics[scale=0.62, angle=270]{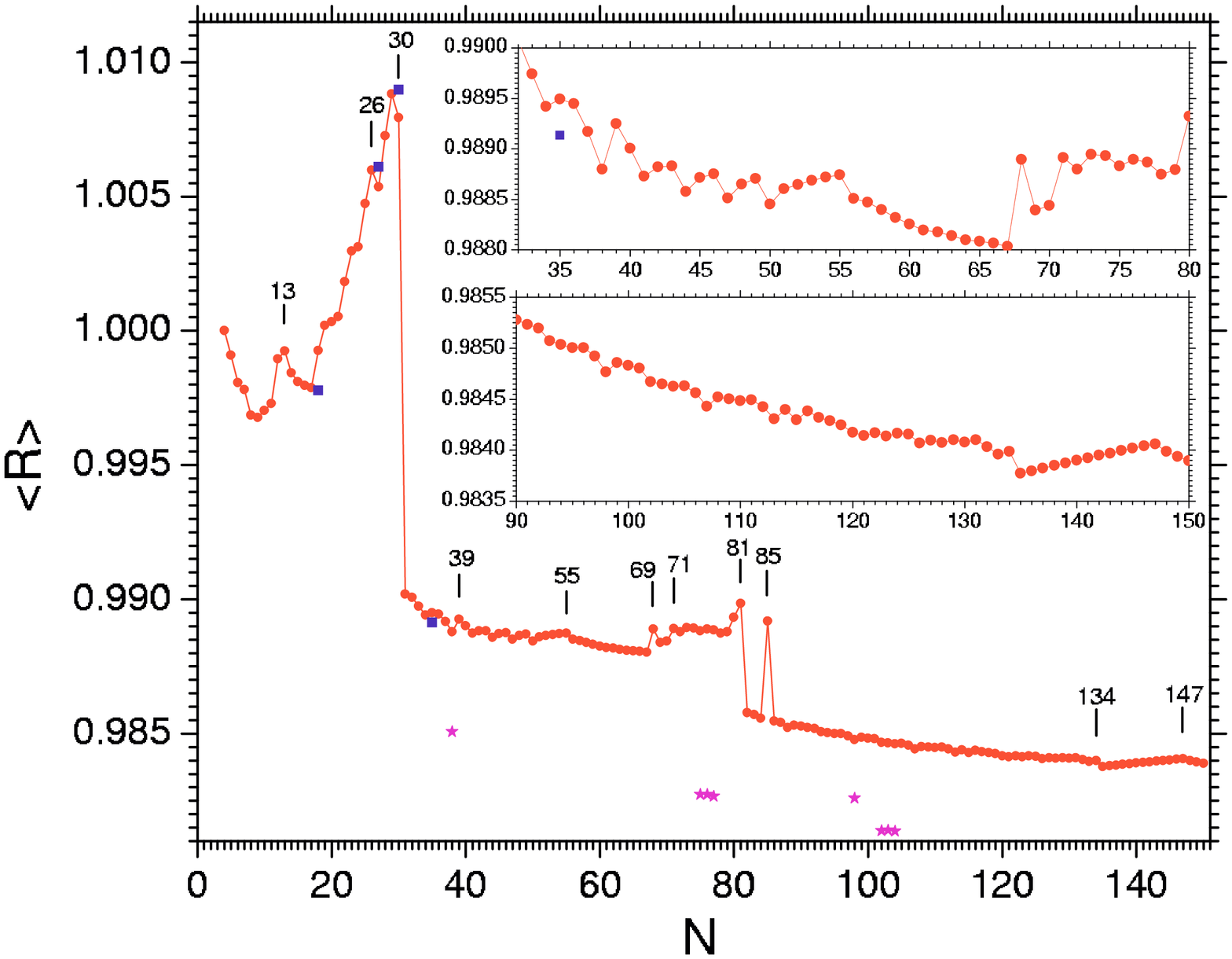}
\end{center}
\caption{
Average bonding length $\left<R\right>$ in the icosahedral LJ global energy
minimum cluster structures as a function of cluster size (circles).
Squares show $\left<R\right>$ for some excited state cluster isomers with the
energy close to the global energy minimum and  stars that for the non-icosahedral
global energy minimum cluster structures.}
\label{R_av}
\end{figure}

To illustrate that the lattice transformations affect significantly
various cluster characteristics, in figures \ref{bond_num} and \ref{R_av},
we present dependences of the average number
of bonds $\left<n\right>$ and the average bonding length
$\left<R\right>$ on cluster size.
These figures demonstrate that for the cluster sizes
$N=31$, $82$, $85$ the dependences
have the step-like irregularities. These irregularities are caused by the
cluster lattice rearrangements.

\begin{figure}
\begin{center}
\includegraphics[scale=0.80]{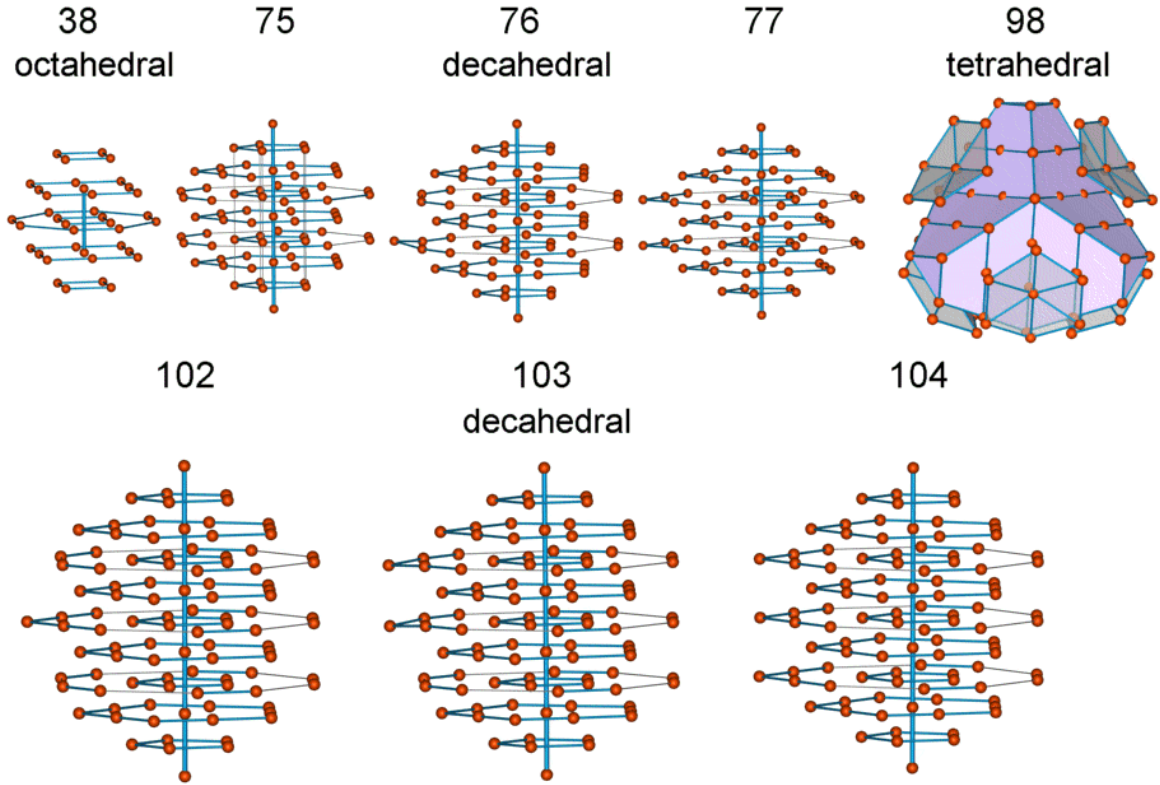}
\end{center}
\caption{
Geometries of non-icosahedral global energy minimum LJ clusters.}
\label{non_icos_globmin}
\end{figure}

In figure \ref{geometry}, we present the geometries of the
global energy minimum
cluster structures with the icosahedral type of atomic packing.
This type of packing is the most favourable for the LJ clusters.
However, other type of packing is also possible and in a few cases
it becomes even more favourable than the icosahedral one.
Within the cluster size range considered, there are only 8 clusters
for which packing other than icosahedral
results in the cluster energy lower
than obtained for the icosahedral clusters. These cluster
structures having the octahedral, decahedral and tetrahedral
type of atomic packing are presented
in figure \ref{non_icos_globmin}. They can not be obtained from
their icosahedral neighbours directly. In order to find these cluster
structures, one needs either to consider a chain of excited state
cluster isomers analogous to the described above for
the formation of the $LJ_{31}$ and $LJ_{82}$ clusters or to treat separately
the growth of the octahedral decahedral and tetrahedral cluster families.
Each of the cluster families should have its own
magic numbers. Only a few of all these cluster configurations can compete
in energy with the clusters based on the icosahedral packing as we
demonstrate it in the next section. Within
the size range considered such situation arises only
for $N=$38, 75-77, 98, 102-104.

Note that in figures \ref{bond_num} and \ref{R_av},
we also show the average number
of bonds $\left<n\right>$ and the average bonding length
$\left<R\right>$ for non-icosahedral global energy minimum cluster isomers and
compare these characteristics with $\left<n\right>$ and
$\left<R\right>$ for clusters with icosahedral
symmetry of the same size.

%
%
%
%
%
%

\subsection{Cluster binding energies}

The binding energy per atom for LJ clusters is defined as follows:

\begin{equation}
E_b/N=-E_N/N 
\label{E_b}
\end{equation}

\noindent
where $E_N$ is the total binding energy of $N$-atomic cluster. In tables
\ref{Energy4_78} and \ref{Energy79_150}, we compile
total energies and point symmetry groups for
the global energy minimum cluster structures within the size range considered. 

\begin{figure}
\begin{center}
\includegraphics[scale=0.73,angle=270]{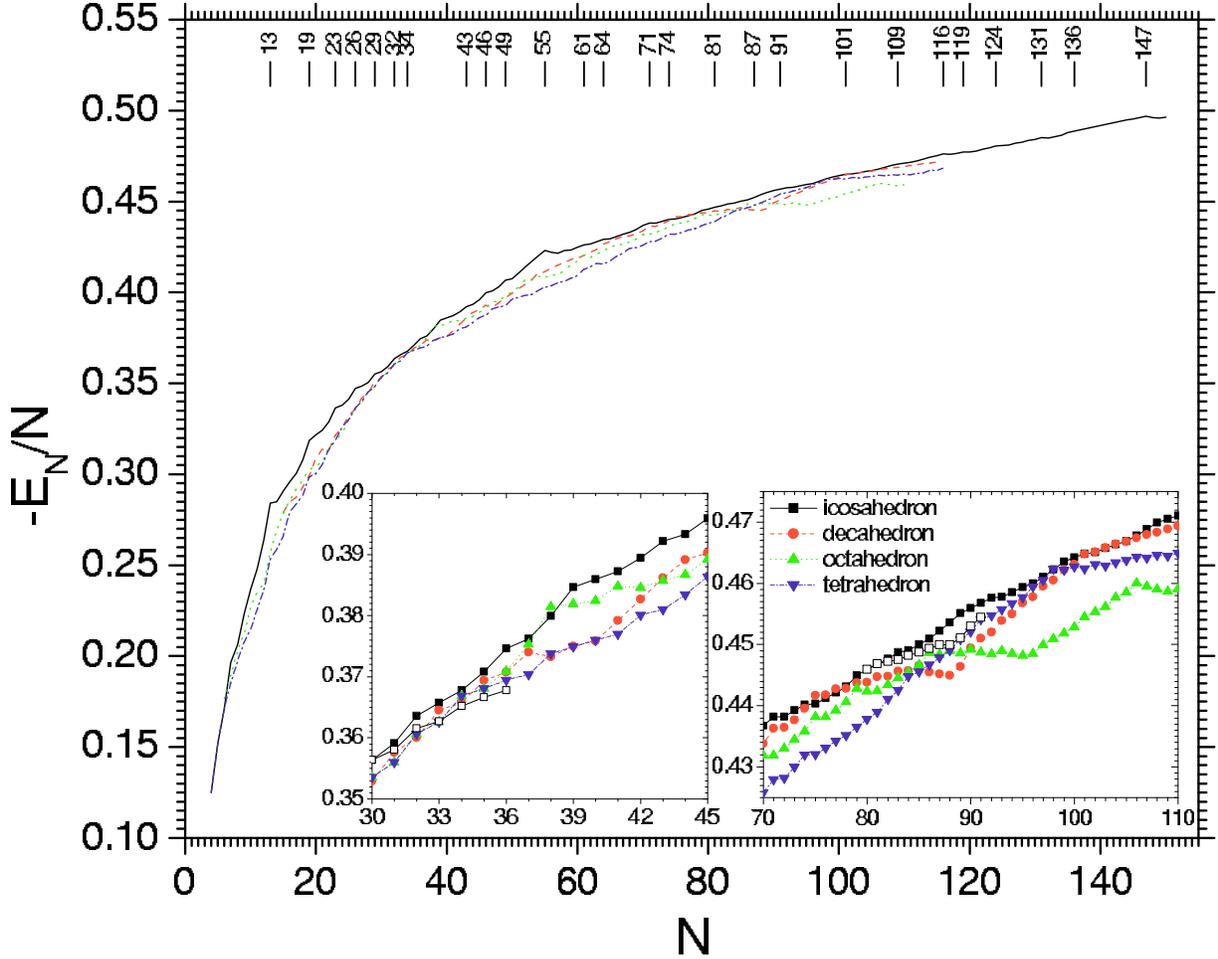}
\end{center}
\caption{
Binding energy per atom for LJ clusters as a function of
cluster size calculated for the cluster chains 
based on the icosahedral (solid line), octahedral
(doted line), tetrahedral
(dash dotted line) and decahedral (dashed line) 
types of packing. In the insets we show the behaviour of the
curves in the regions $30\leq N\leq 45$ and
$70\leq N\leq 110$ in which energies of the clusters with different
type of packing become especially close. We also present the
energies for the chains of clusters growing up from $LJ_{30}$ and
$LJ_{81}$ without rearrangement of their lattice (opened squares).
}
\label{energies_cl}
\end{figure}

Figure \ref{energies_cl} shows the
dependence of the binding energy per atom for LJ clusters
as a function of cluster size.
We have generated the chains of clusters based
on the icosahedral, decahedral, octahedral and tetrahedral types
of lattices with the use of the {\it A3-A5} methods combined
with {\it SE2-SE5} cluster selection criteria.

Figure \ref{energies_cl} shows that the most stable clusters 
are mostly based on the icosahedral type of packing
with exceptions for $N=38$, $75 \leq N \leq 77$, $98$ and
$102 \leq N \leq 104$. In
these cases the octahedral, decahedral, tetrahedral and again decahedral
cluster symmetries respectively become more favourable.
To illustrate this, in the insets to figure \ref{energies_cl},
we show the behaviour of the curves 
in the regions $30\leq N\leq 45$ and $70\leq N\leq 110$ in greater detail.
In  these regions, energies
of the clusters with different type of packing become especially close.

In previous section, we have demonstrated that the structural cluster
properties experience
dramatic change at $N=31$, $82$, $85$. In the contrary to the structural
characteristics, the dependence of binding energy per atom on $N$
behaves smoothly in the vicinity
of these points. In order to stress the role of the cluster lattice
rearrangement in the formation of the global energy minimum cluster structures,
in the inset to figure \ref{energies_cl}, we plot by open squares
the energies for the chains
of clusters growing up from $LJ_{30}$ and $LJ_{81}$ without rearrangement of
their lattice. Clusters in these chains become
more and more energetically unfavourable with the growth cluster size
as compared to the global energy minimum
structures.

On the top of figure \ref{energies_cl} we present the sequence of
magic numbers experimentally measured for noble gases \cite{Echt81,Haberland94}.
It is seen that the magic numbers correspond to the peculiarities in the curve for
the binding energy per atom as a function of cluster size calculated for the
cluster chain based on the icosahedral type of packing. The peculiarities
indicate the enhanced stability of the corresponding clusters.
Note that in experiment 
the magic numbers are seen much better in the regions of $N$,
in which the icosahedrally
packed clusters are energetically the most favourable.
The most prominent peculiarities
arise for the completed icosahedral shells with 13, 55 and 147
atoms. The peculiarities in the dependence of binding energy per atom on $N$ diminish with
the growth of the cluster size due to approaching to the bulk limit,
in which the binding energy per atom is constant. 

\subsection{Liquid drop model}
\label{LD_section}

The main trend of the energy curves 
plotted in figure \ref{energies_cl} can be
understood on the basis of the liquid drop model,
according to which the cluster energy 
is the sum of the volume and the surface energy contributions:

\begin{equation}
E_{N}= - \lambda_V N + \lambda_S N^{2/3} - \lambda_R N^{1/3}
\label{LD_model}
\end{equation}

\begin{figure}
\begin{center}
\includegraphics[scale=0.75]{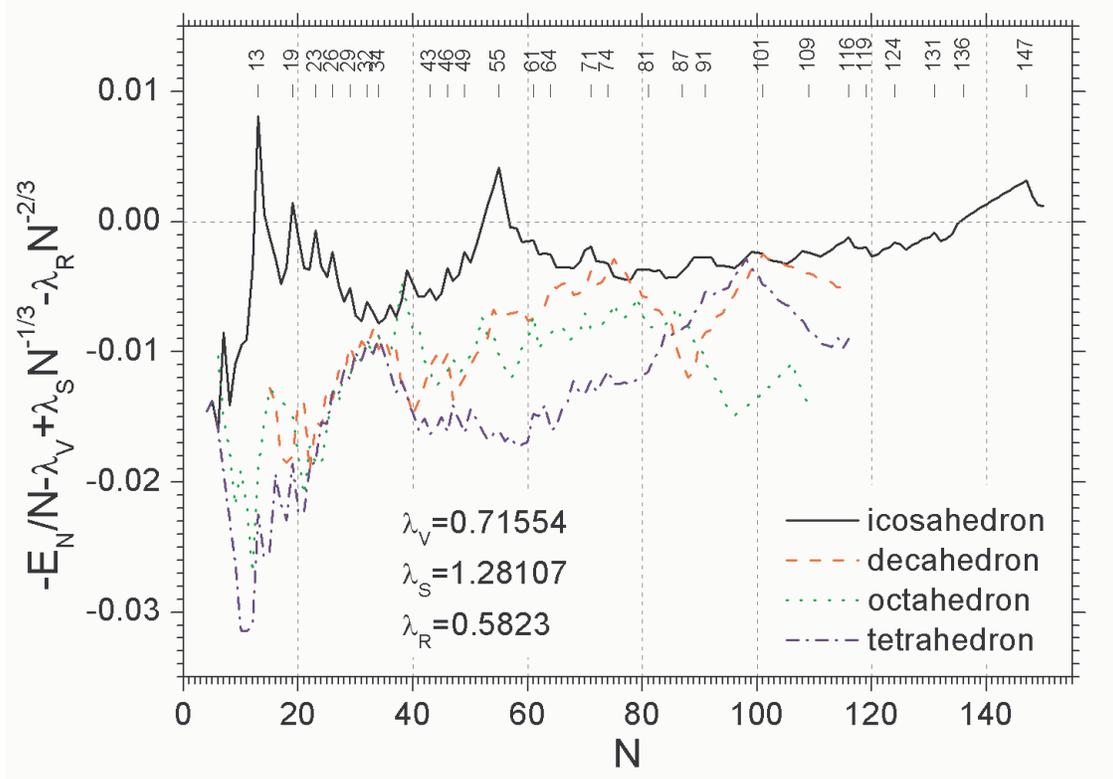}
\end{center}
\caption{Energy curves deviations from the liquid drop
model (\ref{LD_model})
calculated for the cluster chains 
based on the icosahedral (solid line), octahedral
(doted line), tetrahedral
(dash dotted line) and decahedral (dashed line) type of packing.
}
\label{drop_model}
\end{figure}

\noindent
Here the first and the second terms describe 
the volume, and the surface
cluster energy correspondingly. The third term is the
cluster energy arising due to the curvature of the cluster surface.
Choosing constants 
in (\ref{LD_model}) as $\lambda_V=0.71554$,  $\lambda_S=1.28107$
and $\lambda_R=0.5823$, one can
fit the global energy minimum curve plotted in figure \ref{energies_cl}
with the accuracy less than $1 \%$.  The deviations of
the energy curves calculated for
various chains of cluster isomers from the liquid drop
model (\ref{LD_model})
are plotted in figure \ref{drop_model}.
We have fited the energies of the icosahedral clusters in the cluster size range
$4\leq N\leq 150$ using the nonlinear least squares fitting method.

The peaks on these dependences indicate
the increased stability of the corresponding magic clusters.
The ratio between the volume and surface energies
in (\ref{LD_model}) can be characterised by the dimensionless parameter
$\delta=\lambda_V/\lambda_S$, being equal in
our case to $\delta=0.559$. Note, that this value slightly differs
from the previously reported in \cite{GrowProcPRL}, $\delta_{PRL}=0.555$,
which was obtained by fitting
the cluster global energy minima within the same cluster size range.

This result can be compared with those published in \cite{Doye}.
In this paper the energy
of the first four completed icosahedral shells was fitted using equation
(\ref{LD_model}). This fitting gives the value
$\delta=0.576$ differing from our result due to the difference in the fitting
method. In \cite{Doye} the energies
of the 4 selected clusters have been used, while we fit the energies of all
the clusters within the size range $4\leq N\leq 150$

Note that a similar model is used in nuclear physics for the description
of the nuclei binding energy.
For the nuclear matter, the constant $\delta$ is equal to $\delta=0.903$  \cite{BM}.

\subsection{Cluster magic numbers}

\begin{figure}
\begin{center}
\includegraphics[scale=0.67]{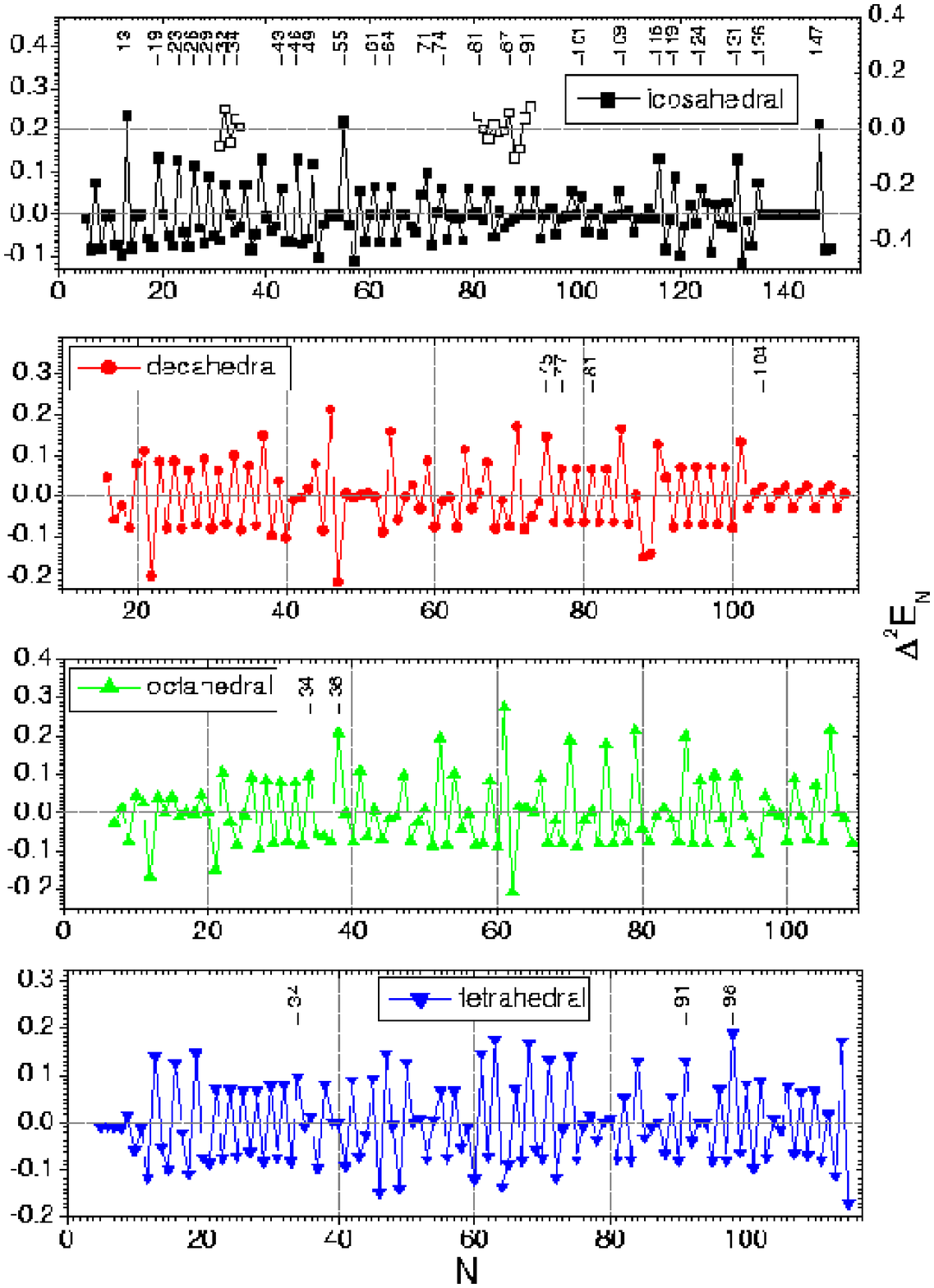}
\end{center}
\caption{Second derivatives $\Delta^2 E_N$ for
the icosahedral (squares),
decahedral (circles), octahedral (upper triangles) and tetrahedral
(lower triangles) cluster isomer chains. Open squares show
$\Delta^2 E_N$ for the chains of energetically unfavourable
clusters growing from $LJ_{30}$ and $LJ_{81}$ without rearrangement
of their lattice.
}
\label{magic_numbers}
\end{figure}

The dependence of the binding energies per atom for the most stable cluster
configurations on $N$ allows one to generate the sequence of the
cluster magic numbers. In figure \ref{magic_numbers}, we plot the second
derivatives $\Delta^2 E_N$ for the cluster chains with icosahedral,
decahedral, octahedral and tetrahedral type of packing.

We compare the obtained dependences with the experimentally measured abundance 
mass spectra
for the noble gas clusters \cite{Echt81,Haberland94}
(see figure \ref{mass_spectra}) and
establish the striking correspondence
of the peaks in the experimentally measured mass spectra with those
in the $\Delta^2 E_N$ dependence calculated for the icosahedral type of
clusters. The sequence of the magic numbers
for the Ar, Xe and Kr clusters reads: 
13, 19, 23, 26, 29, 32, 34, 43, 46, 49,
55, 61, 64, 71, 74, 81, 87,  91,  101, 109, 116, 119, 124, 131, 136, 147
\cite{Echt81,Haberland94}.
The most prominent peaks in this sequence 13, 55 and 147 
correspond to the closed icosahedral
shells, while other numbers correspond to the filling of
various parts of the icosahedral shell. 

The connection between the second derivatives $\Delta^2 E_N$
and the peaks in the abundance mass spectrum of clusters
one can understand using the following simple model.
Let us assume that the mass spectrum of clusters is formed
in the evaporation process. This means that changing
the number of clusters $n_N$ of the size $N$ in the
cluster ensemble takes place due to the evaporation of an atom by the clusters
of the size $N$ and $N+1$, i.e.

\begin{equation}
\Delta n_N \sim n_{N+1} W_{N+1 \rightarrow N} - n_{N} W_{N \rightarrow N-1}
\label{Delta_evaporation}
\end{equation}

\noindent
where the evaporation probabilities  are proportional to

\begin{eqnarray}
W_{N+1 \rightarrow N} &\sim& e^{-\frac{E_N+E_1-E_{N+1}}{kT}}\\
\nonumber
W_{N \rightarrow N-1} &\sim& e^{-\frac{E_{N-1}+E_1-E_N}{kT}}\\
\nonumber
\label{evaporation_prob}
\end{eqnarray}

\noindent
Here $T$ is the cluster temperature, $k$ is the Bolzmann constant.
In our model $E_1=0$, so after simple transformations the equation for
$\Delta n_N$ reads as:

\begin{eqnarray}
\Delta n_N &\sim& n_{N+1} e^{-\frac{E_{N}-E_{N+1}}{kT}}
(1-\frac{n_{N}}{n_{N+1}} e^{-\frac{E_{N-1}-2E_{N}+E_{N+1}}{kT}})\\
\nonumber
&\sim&n_{N} e^{-\frac{E_{N}-E_{N+1}}{kT}}
(1-e^{-\frac{E_{N-1}-2E_{N}+E_{N+1}}{kT}})
\nonumber
\label{Delta_evap}
\end{eqnarray}

\noindent
Here we assumed that $n_{N+1} \sim n_N \sim n_{N-1} \gg 1$. 
Let us now estimate the relative abundances in the mass spectrum
for Ar clusters for temperatures about 100 K.
The exponent $e^{-\frac{E_{N}-E_{N+1}}{kT}}$ influences the absolute
value of $\Delta n_N$. This factor becomes exponentially small at
$kT \ll E_{N}-E_{N+1}$, which for the Ar clusters means $T \ll 800 K$,
because $\left< \Delta E_{N}^{Ar} \right> = 0.071 eV \sim 800 K$. The small
value of this factor results in the growth of the characteristic period
of the evolution of $n_N$ with time. The factor in the brackets determines the
relative cluster abundances. Indeed, its positive value for certain $N$ leads
to the growth of the corresponding clusters in the system, while the
negative value of the factor to the opposite behaviour. The factor in the
brackets is characterised by $\Delta^2 E_{N}=E_{N-1}-2E_{N}+E_{N+1}$,
which is for the Ar clusters
$\left< |\Delta^2 E_N| \right> =0.008 eV \sim 90 K$.
Thus, for temperatures $T \agt 90 K$ the exponent in the brackets can be
expanded. In this case one derives

\begin{equation}
\Delta n_N \sim n_N e^{-\frac{E_{N}-E_{N+1}}{kT}} \frac{\Delta^2 E_N}{kT}
\label{Delta_evap_fin}
\end{equation}

\noindent
This equation demonstrates, that
positive values of $\Delta^2 E_N$ leads to the enhanced abundance
of the corresponding clusters.

For Xe and Kr clusters this approximation becomes applicable at somewhat
larger $T$, because the depth of the LJ potential, $\varepsilon$,
for this type of clusters, is larger than
for Ar clusters. For Ar clusters $\varepsilon=12.3$ meV while for
Kr and Xe it is $17.2$ and $24.3$ meV respectively.

Within the size range considered there are only five magic numbers, which
can not be directly explained on the basis of $\Delta^2E_N$ calculated
for the chain of icosahedral clusters.
This situation takes place for $N=$34, 81, 87, 91 and 136.
However, the origin of these magic numbers can also be understood.

The magic number $N=$34 is not seen in the chain of the energetically
most favourable icosahedral clusters,
because of the lattice rearrangement that occurs for  $N=$31.
Beginning from $LJ_{31}$ a new cluster growing chain becomes energetically more
favourable, but
the chain of clusters growing from $LJ_{30}$ without its lattice rearrangement
turns out to be rather close to the global
energy minimum  chain (see figure \ref{energies_cl}).
As a result, in the vicinity of the transition point both chains
influence the magic number formation. Thus, for $N=$34,
the positive peak in the $\Delta^2E_N$ for the chain
of the global minimum clusters is absent, but it for
present in the chain of clusters growing
from the $LJ_{30}$ cluster without its lattice rearrangement
(see open squares in figure \ref{magic_numbers}).
This magic number can also be seen
for the octahedral and tetrahedral cluster chains,
which come energetically rather close to the icosahedral chain in this region of $N$.

For the $LJ_{81}$ cluster the situation is similar. Here again a
structural rearrangement of the cluster lattice 
takes place influencing the magic number formation. 
Thus, the magic numbers 81, 87, 91 arise in the chain of clusters
growing from the $LJ_{81}$ cluster without its lattice rearrangement.
The number 81 arises also in the chain of the decahedral clusters,
which in this region of $N$ is energetically
close to the global energy minimum (see figure \ref{magic_numbers}).
The chain of tetrahedral clusters affects
the formation of the 91 magic number due to the same reason.
Though this magic number is absent in the sequence of magic numbers
generated for global energy
minimum cluster structures, it is well pronounced in the
magic number sequence for the tetrahedral clusters and the chain of
clusters growing from $LJ_{81}$ without its lattice rearrangement.

Another magic number, which is masked in the magic number sequence for
the icosahedral clusters is $N=$136. This number is masked
because of the radical surface rearrangement of atoms needed
for obtaining its configuration from the $LJ_{135}$ one. The $LJ_{135}$ cluster
is the truncated 147 icosahedron. It possesses the $I_h$ point symmetry group.
It can be obtained
from the $LJ_{134}$ cluster by a rearrangement of 7
atoms located on the cluster surface.
This cluster configuration can be obtained via a long chain of excited state
cluster configurations which starts from the $LJ_{55}$ icosahedral cluster.
In experiment, the formation of the global energy minimum $LJ_{135}$ cluster
structure should occur with relatively low probability because of
the reasons outlined above.
Instead, it is feasible to assume that for this particular cluster size different
cluster isomers with energies close to the global minimum can be readily created.
This influences the 135 magic number formation
resulting in its shift to $N=136$.

\begin{figure}
\begin{center}
\includegraphics[scale=0.90]{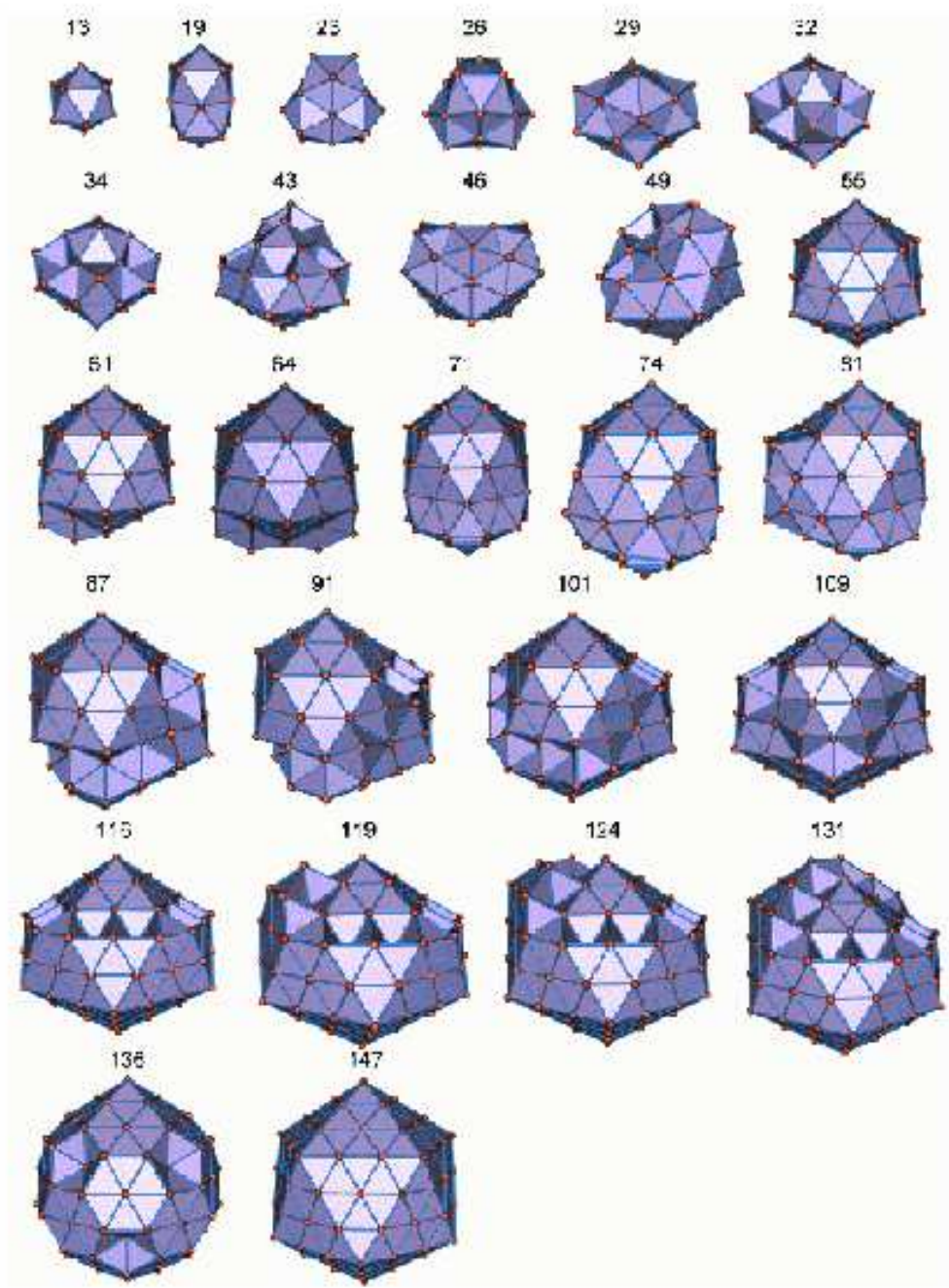}
\end{center}
\caption{
Geometries of the magic LJ clusters. 
The labels indicate the cluster size.
}
\label{cl_images}
\end{figure}

Note that $\Delta^2 E_N$ values calculated for the chain of icosahedral
clusters in the size range $136\leq N\leq 146$ are very close to zero.
This fact can be explained as follows. There are 12 
vacancies in the outer icosahedral shell of the $LJ_{135}$ cluster,
which are located at the vertices of
the $LJ_{147}$ icosahedron. These vacancies are sequentially filled
within the size range $136\leq N\leq 146$.
Within this size range the energy difference between the two neighbouring clusters
practically  does not vary and is determined by the energy of an atom placed in one
of the icosahedron vertice vacancies. Such a situation takes place,
because the distances between 
any of two vertices are much
larger than the pairing bond length in the LJ potential and the
interaction between the atoms placed in the icosahedron vertices is neglible.
As a result, of this the total energy grows nearly linearly and
$\Delta^2 E_N \approx 0$

In figure \ref{cl_images}, we plot images of the magic global
energy minimum cluster structures.  
In this figure it is clearly seen how one icosahedral shell emerges
from another. This figure demonstrates that the process of the
$LJ_{55}$ icosahedron cluster formation has
similarities with the formation of the $LJ_{147}$
cluster. In both cases, at the first stage
a cap is formed on the top of the completed icosahedron shell,
see the $LJ_{19}$ and $LJ_{71}$ magic clusters.
Then, in both cases a structural rearrangement of the cluster
takes place. When the rearrangement of the cluster lattice is
completed ($LJ_{31}$, $LJ_{82}$), the growth of the cluster occurs by
filling the atomic vacancies in the vicinity of the cluster surface
up to the point when a new icosahedral shell is formed.

We have demonstrated that the magic numbers for the icosahedral clusters
map well enough the experimentally
observed magic numbers.
In figure \ref{non_icos_globmin}, we present the images of the
non icosahedral global energy minimum cluster structures.
Experimentally these clusters are not found to be 
the magic clusters, although, they are
the global minimum clusters being magic for
the chains of clusters based on the decahedral, octahedral or tetrahedral
type of packing (see figures \ref{energies_cl} and \ref{magic_numbers}). 
This fact can be understood if one takes into account
that the chains of clusters with different type of atomic
packing are formed independently and the transition of clusters
from one chain to another at certain $N$ occurs with the small probability.
From the
binding energy analysis, it is clear that the chain of icosahedral clusters
prevails. Thus, in experiment, the icosahedral clusters
mask the clusters belonging to the
chains of non-icosahedraly packed clusters, even when the
non-icosahedral cluster structures happen to become energetically more favourable.

%

\subsection{Spontaneous cluster fusion}

Using the {\it A2} method and the {\it SE2} cluster selection
criterion we have generated a chain of energetically unfavourable
cluster isomers with $N \leq 84$ as an example of the spontaneous cluster
growth.

\begin{figure}
\begin{center}
\includegraphics[scale=0.62,angle=270]{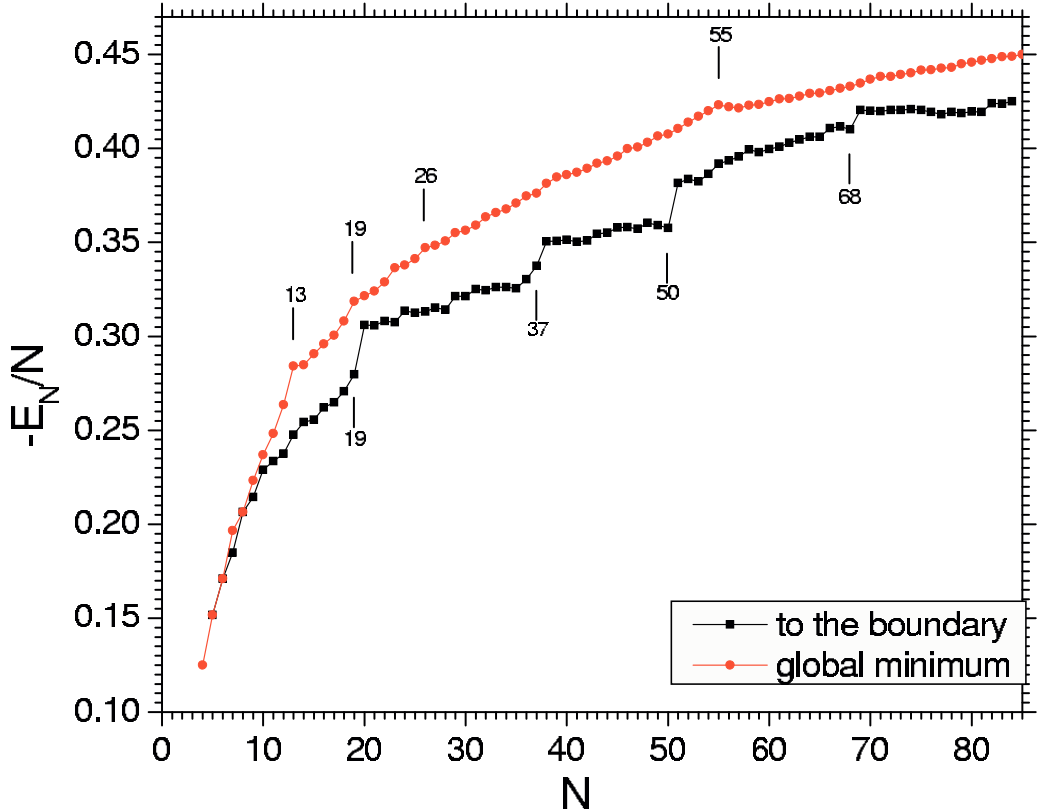}
\end{center}
\caption{
Binding energies per atom for a chain of spontaneously produced clusters 
in comparison with the binding energies per atom for the global energy minimum
cluster structures.
}
\label{spont_energy}
\end{figure}

\begin{figure}
\begin{center}
\includegraphics[scale=0.60]{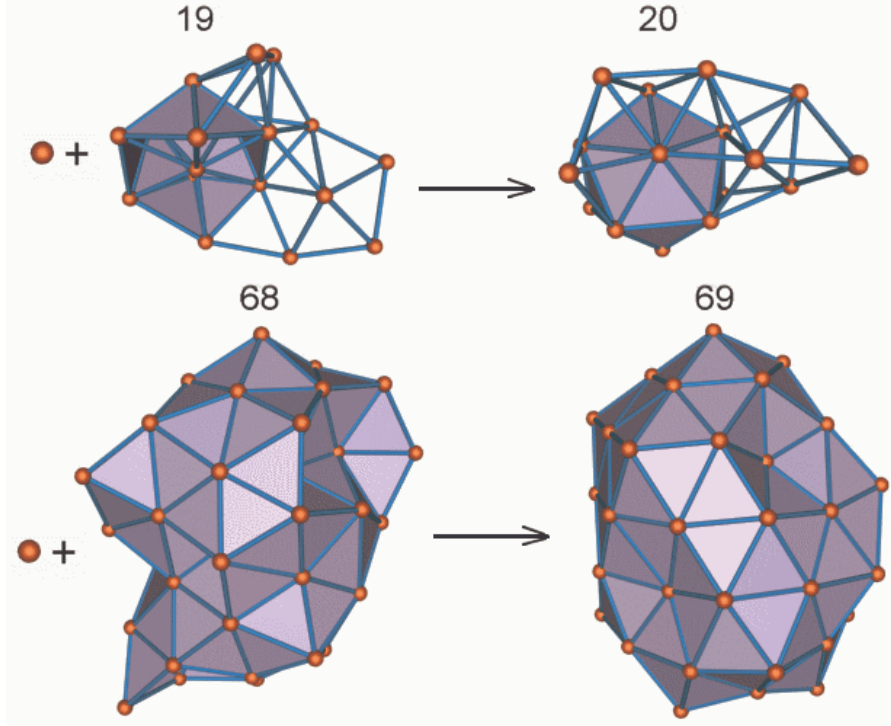}
\end{center}
\caption{
Structural rearrangements in spontaneously generated cluster structures
with $N=19-20$ and $N=68-69$. Fusion of just one atom to the cluster
boundary rearranges its structure and leads to the formation of a stable
cluster core.
}
\label{struct_rear_spont}
\end{figure}

In figure \ref{spont_energy}, we compare the binding energies per atom
for the chain of the spontaneously produced clusters with the binding
energies per atom calculated for the
global energy minimum cluster structures.
This figure shows that the binding energy per atom
for the spontaneously generated cluster structures are systematically lower than
those for the global energy minimum structures.
However, the two dependences behave quite differently.
Thus, at certain cluster sizes, the binding energy per atom for the spontaneously
generated chain of clusters increases in a step-like manner. In the size range
considered, this occurs for the cluster sizes
$N=19-20,$ $37-38$, $50-51$, $68-69$, $81-82$. These irregularities
originate from the significant structural rearrangement taking
place in the spontaneously generated cluster structures.
In figure \ref{struct_rear_spont}, we illustrate such a rearrangement
for the clusters with $N=19-20$ and $N=68-69$.

The structural rearrangement in the $LJ_{19}$
cluster leads to the formation of the
compact 13-atomic icosahedron shell. Indeed, in
the initial $LJ_{19}$ cluster isomer, only a part of the
icosahedron can be identified (see
figure \ref{struct_rear_spont}). Fusion of just one atom to the cluster boundary
rearranges its structure and forms the compact icosahedral shell as a cluster
core. This transformation stabilizes the cluster structure and makes it
energetically much more favourable.

The situation with the $LJ_{68}$ cluster is similar. Again, we consider
a highly asymmetric cluster
isomer (see figure \ref{struct_rear_spont}), which
deviates significantly from the global energy minimum cluster structure
based on the 55-atomic icosahedron shell (see figure \ref{geometry}).
Fusion of a single atom to the boundary of the initial cluster
configuration rearranges its
structure completely forming a relatively compact $LJ_{69}$ cluster isomer
(see figure \ref{struct_rear_spont}). The $LJ_{69}$ cluster configuration
possesses the evident
elements of the 55-atomic icosahedron shell, which are absent in the 
initial $LJ_{68}$ cluster isomer.
The compact configuration of the found $LJ_{69}$ cluster
isomer brings its binding energy per atom  close to the binding energy per
atom calculated for the global energy minimum cluster structure.

\subsection{Symmetrical cluster fusion}

\begin{figure}
\begin{center}
\includegraphics[scale=0.73,angle=270]{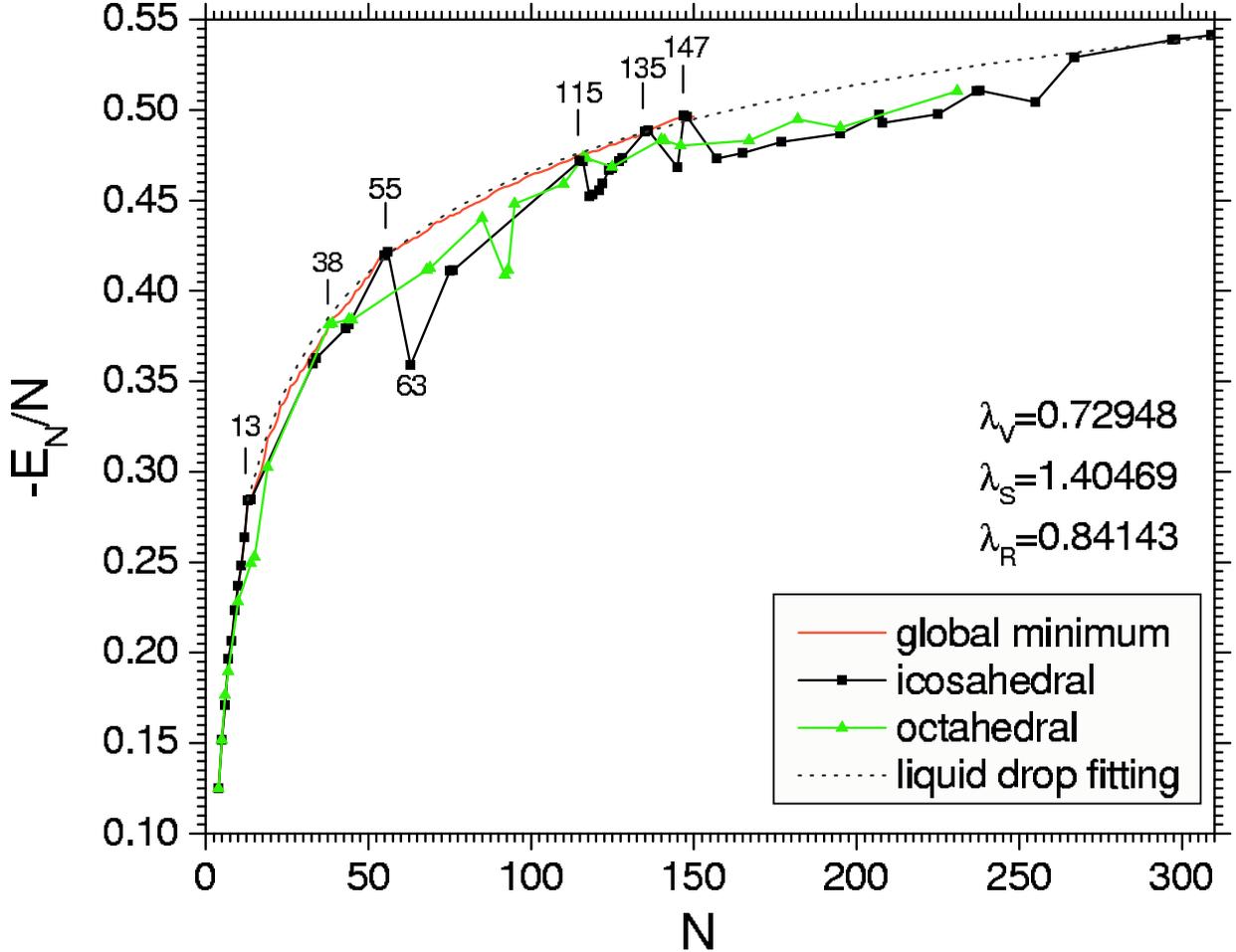}
\end{center}
\caption{
Binding energies per atom for sequences of symmetrical clusters with icosahedral (squares)
and octahedral (triangles) types of packing and their
comparison with the binding energies per atom for the global energy minimum
cluster structures (solid line) and the liquid drop model (\ref{LD_model})
fitting (dotted line). 
}
\label{symm_energy}
\end{figure}

Using the {\it SY1-SY3} methods, we now consider the formation of symmetrical
clusters and perform their energy analysis. We have generated the cluster configurations
with $N\leq 309$ possessing the icosahedral and octahedral symmetry.

In figure \ref{symm_energy}, we plot the binding energies per atom calculated for
clusters possessing the icosahedral and octahedral symmetry
and compare them with the binding energies per atom calculated for the global energy
minimum cluster structures. Figure \ref{symm_energy} demonstrates the irregular
behaviour of the binding energies of the calculated symmetrical cluster
configurations. Note that this irregularities are not of the physical
nature, because the {\it SY1-SY3} methods
do not model any concrete physical scenario of the cluster fusion process.
These methods rather represent an efficient mathematical algorithm for the generation of
symmetrical cluster configurations (see section \ref{model} for details).

It is interesting that some of the found symmetrical cluster
configurations within the size range $N\leq 150$ appear to
be the global energy minimum cluster structures. This is the case
for the clusters with $N=$13, 55, 135, 147, possessing
the icosahedral point symmetry group, $I_H$. The binding energy of the highly
symmetrical cluster isomer $LJ_{115}$ possessing also the $I_H$ point symmetry group
is only about $0.0033$ units smaller than
the energy of the corresponding global energy minimum.
It is interesting that the $LJ_{115}$ global energy minimum
cluster structure possesses the $C_{5V}$ point symmetry group, being lower than
the $I_H$ point symmetry group.
The fact that the cluster isomer with the lower symmetry becomes energetically
more favourable can be understood after counting the total
number of bonds in the system (see section
\ref{fusion} for details). In the $LJ_{115}$
icosahedral isomer the total number of bonds is equal to 524, while
in the global energy minimum isomer it is equal to 534.
Here we have used the cut-off value equal to $d=1.2\cdot 2^{1/6}\sigma$.
Comparison of the two number shows that the large number of bonds can be
created in a system with the lower symmetry and that the larger number
of bonds means the higher stability of the system.

Among the octahedral cluster configurations found with the use of the same technique
there are also a few ones with the energy being very close to
the cluster global energy minimum.
Thus, the truncated octahedron $LJ_{38}$ appears to be the only global
energy minimum cluster structure within the size range of $N\leq 150$
possessing the octahedral symmetry (see section \ref{fusion} for
more details). The $LJ_{116}$ cluster isomer has also the form of
truncated octahedron. Its binding energy is smaller than the binding energy
of the global energy minimum cluster structure of the same size
on $0.0024$ unit.

\begin{figure}
\begin{center}
\includegraphics[scale=0.72]{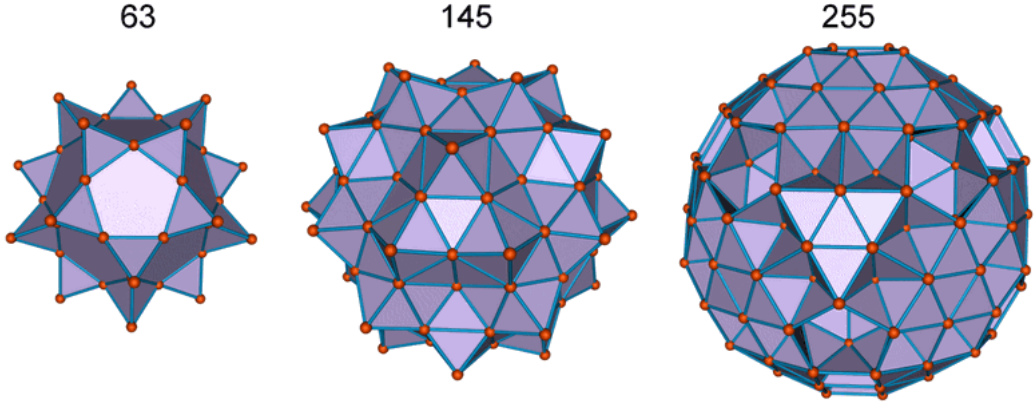}
\end{center}
\caption{
Clusters with icosahedral point symmetry group having low binding energy per atom.
}
\label{low_en_isomers}
\end{figure}

Note that some highly symmetric clusters,
like the icosahedral $LJ_{63}$, $LJ_{145}$, $LJ_{255}$
can have relatively low binding energy as compared
to their neighbours (see figure \ref{symm_energy}).
The geometries of these clusters are shown in figure \ref{low_en_isomers}.
They appear to be relatively low stable, because the total
number of bonds in these clusters
(222, 654 and 1194 respectively) is significantly lower than
it is in the corresponding global energy minimum cluster
structures (269 for $LJ_{63}$, 684 for $LJ_{145}$, for $LJ_{255}$
the total number of bonds in the global energy minimum is not known).

The calculation of the symmetrical cluster configuration and their energies turns
out to be very useful for the preliminary
analysis of the binding energy curve in a wide
range. Thus, fitting the points $N=13,55,115,135,147,267,297,309$
with use of the liquid drop model equation (\ref{LD_model}),
one derives the following values of $\lambda_V=0.72948$, $\lambda_S=1.40469$,
$\lambda_R=0.84143$, which are in a good agreement with the results
following from the global
energy minimum calculations, see section \ref{LD_section}. We note that the performed
analysis of symmetrical cluster configurations is essentially
easier than the complete
analysis of the global energy minima and thus it can be used for
the efficient calculation of cluster binding energies in a wide range of $N$
at relatively high level of accuracy.

\section{Conclusion}
\label{conclusion}

In this paper we have discussed the classical models for
the cluster growing process, but  our ideas 
can be easily generalized on the quantum case and
be applied to the cluster systems with different than LJ type 
of the inter-atomic interaction. It would be interesting
to see to which extent the parameters of  inter-atomic interaction
can influence the cluster growing process and the corresponding
sequence of  magic numbers or whether the crystallization
in the nuclear matter consisting of alpha particles and/or nucleons 
is possible. Studying cluster thermodynamic
characteristics with the use of the developed
technique is another interesting problem which is left open for
future considerations.

\section{Acknowledgments}

The authors acknowledge support of this work by the Studienstiftung des deutschen Volkes,
Alexander von Humbolt Foundation, DFG, Russian Foundation of
Basic Research, Russian Academy of Sciences and the Royal Society of London.

\appendix

\section{Algorithm details}
\label{algorithms}

In this section we provide some details of the algorithms used
in the computations \cite{NanoMeeting}.

In the described algorithm, the potential energy 
of selected atom decreases during its motion.  
Therefore, the total potential energy of the cluster decreases 
at each step of the algorithm and it finally converges to a stable 
cluster configuration. 
In most cases, this convergence is rather fast. 
The time required for the obtaining of a stable configuration with the 
error $\Delta E$ of the potential energy is usually proportional to 
$-ln \Delta E$. 
To speed up the process of conversion the procedure for the 
determination of the LJ force acting on the atom 
was written in Assembler x86. The performance of the calculations increased 
approximately 1.5 times comparing to the version written entirely in C.

Note that the described algorithm of the kinetic energy absorption
is much more efficient as compared to those reducing the kinetic
energy of moving atoms at each step of the Runge-Kutta 
integration procedure. 
Firstly, the following method was used: during the motion on each 
Runge-Kutta step the speed of an atom was decreased $k$ times. 
Therefore, the whole energy of oscillations that occur near the 
equilibrium position decreases because of the decreasing of the kinetic 
energy. So far, when the speed becomes very small, 
atoms start to oscillate very close to the potential energy minimum 
position. This method is rather simple, but too much time is
required for atom coordinates to converge to their equilibrium
position. Therefore we introduced another sufficient energy
absorption algorithm, that was described in section \ref{model}.

To prevent the fragmentation of the cluster caused by the unfavourable
initial conditions, the adaptive choice of time-step in
the Runge-Kutta integration procedure has been implemented.
The fragmentation of the system might happen, for example, in the case when
initially a pair of atoms are placed at a small distance one from another 
so that they experience a very strong repulsive force. In such
a situation the moving atom gains high acceleration and during the
first step of the Runge-Kutta procedure might go far away from the cluster.
If the force acting on this atom in its final position 
is smaller than a given value $F_{min}$,
it will be assumed in equilibrium and 
it will never be attracted back to the cluster, although 
the final configuration won't be stable. 
The adaptive choice of the integration time-step prevents the atoms
to fly apart. The idea of this procedure is rather simple.
Each time step $\Delta t$ in the Runge-Kutta integration procedure is
repeated with the time step $\Delta t/2$. If the coordinate or
velocity  change on the interval $\Delta t/2$ 
is found too large then the integration step $\Delta t$ is reduced 
and the procedure repeats. If the coordinate 
change is too small than the larger value of the interval $\Delta t$
is set and the calculation is repeated again.
The adaptive choice of the integration time step prevent the
atoms to fly away, because in the unfavourable initial
conditions situation
the integration time-steps will become small, though the velocity 
will be high. Nevertheless, when the atom reaches the boundary of 
the cluster, it will stop: at that moment its velocity starts 
to decrease because of the attraction to the cluster.

\section{Tables}
\label{tables}

\begin{table*}[hp]
\caption{Total energies and point symmetry groups
for LJ global energy minimum isomers within the size range $4\leq N\leq78.$}
\label{Energy4_78}
\begin{ruledtabular}
\begin{tabular}{ccccccccc}
 
     \multicolumn{1}{c}{N } &
     \multicolumn{1}{c}{Point group} &
     \multicolumn{1}{c}{Energy} &
     \multicolumn{1}{c}{N } &
     \multicolumn{1}{c}{Point group} &
     \multicolumn{1}{c}{Energy}&
     \multicolumn{1}{c}{N } &
     \multicolumn{1}{c}{Point group} &
     \multicolumn{1}{c}{Energy}\\
\\
\hline 

4 & $T_{D} $ & -0.500000  &29& $D_{3H}$ & -10.29895  &54& $C_{5V}$ & -22.68405\\                           
5 & $D_{3H}$ & -0.758654  &30& $C_{2V}$ & -10.69055  &55& $I_{H} $ & -23.27071\\                           
6 & $O_{H} $ & -1.059339  &31& $C_{S} $ & -11.13220  &56& $C_{3V}$ & -23.63693\\                           
7 & $D_{5H}$ & -1.375449  &32& $C_{2V}$ & -11.63629  &57& $C_{S} $ & -24.02855\\                           
8 & $C_{S} $ & -1.651791  &33& $C_{S} $ & -12.07023  &58& $C_{3V}$ & -24.53151\\      
9 & $C_{2V}$ & -2.009447  &34& $C_{2V}$ & -12.50371  &59& $C_{2V}$ & -24.97817\\      
10& $C_{3V}$ & -2.368544  &35& $C_{1} $ & -12.97972  &60& $C_{S} $ & -25.48962\\      
11& $C_{2V}$ & -2.730498  &36& $C_{S} $ & -13.48545  &61& $C_{2V}$ & -26.00074\\      
12& $C_{5V}$ & -3.163967  &37& $C_{1} $ & -13.91947  &62& $C_{S} $ & -26.44616\\      
13& $I_{H} $ & -3.693900  &38& $O_{H} $ & -14.49404  &63& $C_{1} $ & -26.95748\\      
14& $C_{3V}$ & -3.987096  &39& $C_{5V}$ & -15.00277  &64& $C_{S} $ & -27.46835\\      
15& $C_{2V}$ & -4.360219  &40& $C_{S} $ & -15.43749  &65& $C_{2} $ & -27.91429\\      
16& $C_{S} $ & -4.734645  &41& $C_{S} $ & -15.87802  &66& $C_{1} $ & -28.42588\\      
17& $C_{2} $ & -5.109833  &42& $C_{S} $ & -16.35646  &67& $C_{S} $ & -28.93767\\      
18& $C_{5V}$ & -5.544246  &43& $C_{S} $ & -16.86372  &68& $C_{1} $ & -29.44955\\       
19& $D_{5H}$ & -6.054982  &44& $C_{1} $ & -17.30739  &69& $C_{5V}$ & -29.99021\\       
20& $C_{2V}$ & -6.431420  &45& $C_{1} $ & -17.81541  &70& $C_{5V}$ & -30.57435\\       
21& $C_{2V}$ & -6.807048  &46& $C_{2V}$ & -18.39003  &71& $C_{5V}$ & -31.11247\\                            
22& $C_{S} $ & -7.234149  &47& $C_{1} $ & -18.83435  &72& $C_{S} $ & -31.55310\\       
23& $D_{3H}$ & -7.737039  &48& $C_{S} $ & -19.34996  &73& $C_{S} $ & -32.06578\\       
24& $C_{S} $ & -8.112401  &49& $C_{3V}$ & -19.92432  &74& $C_{S} $ & -32.57571\\       
25& $C_{S} $ & -8.531055  &50& $C_{S} $ & -20.37916  &75& $D_{5H}$ & -33.12436\\       
26& $T_{D} $ & -9.026301  &51& $C_{2V}$ & -20.93783  &76& $C_{S} $ & -33.57457\\                            
27& $C_{2V}$ & -9.406132  &52& $C_{3V}$ & -21.51917  &77& $C_{2V}$ & -34.09029\\       
28& $C_{S} $ & -9.818533  &53& $C_{2V}$ & -22.10025  &78& $C_{S} $ & -34.56620\\       
                                                                                      
\end{tabular}                   
\end{ruledtabular}              
\end{table*}

\begin{table*}[hp]
\caption{The same as table \ref{Energy4_78} but
within the size range $79\leq N\leq 150$.}
\label{Energy79_150}
\begin{ruledtabular}
\begin{tabular}{ccccccccc}
 
     \multicolumn{1}{c}{N } &
     \multicolumn{1}{c}{Point group} &
     \multicolumn{1}{c}{Energy} &
     \multicolumn{1}{c}{N } &
     \multicolumn{1}{c}{Point group} &
     \multicolumn{1}{c}{Energy}&
     \multicolumn{1}{c}{N } &
     \multicolumn{1}{c}{Point group} &
     \multicolumn{1}{c}{Energy}\\
\\
\hline 

 79& $C_{2V}$ & -35.15091  &104& $C_{2V}$ & -48.50722  &129& $C_{S} $ & -62.37172\\        
 80& $C_{S} $ & -35.67363  &105& $C_{1} $ & -49.02221  &130& $C_{1} $ & -62.93926\\        
 81& $C_{2V}$ & -36.19530  &106& $C_{1} $ & -49.58842  &131& $C_{2V}$ & -63.53680\\        
 82& $C_{1} $ & -36.71254  &107& $C_{S} $ & -50.16726  &132& $C_{1} $ & -64.00352\\        
 83& $C_{2V}$ & -37.24367  &108& $C_{S} $ & -50.75275  &133& $C_{S} $ & -64.58527\\        
 84& $C_{1} $ & -37.72143  &109& $C_{1} $ & -51.28426  &134& $C_{3V}$ & -65.18385\\         
 85& $C_{3V}$ & -38.25465  &110& $C_{S} $ & -51.81566  &135& $I_{H} $ & -65.85651\\         
 86& $C_{1} $ & -38.78204  &111& $C_{S} $ & -52.33903  &136& $C_{5V}$ & -66.45444\\         
 87& $C_{S} $ & -39.34151  &112& $C_{S} $ & -52.90622  &137& $C_{2V}$ & -67.05262\\         
 88& $C_{S} $ & -39.91939  &113& $C_{S} $ & -53.48289  &138& $C_{3V}$ & -67.65107\\         
 89& $C_{3V}$ & -40.50449  &114& $C_{S} $ & -54.06942  &139& $C_{2V}$ & -68.24949\\         
 90& $C_{S} $ & -41.03616  &115& $C_{5V}$ & -54.64636  &140& $C_{S} $ & -68.84789\\         
 91& $C_{S} $ & -41.56759  &116& $C_{5V}$ & -55.23411  &141& $C_{5V}$ & -69.44655\\         
 92& $C_{3V}$ & -42.09876  &117& $C_{1} $ & -55.69023  &142& $C_{S} $ & -70.04488\\         
 93& $C_{1} $ & -42.57314  &118& $C_{S} $ & -56.23080  &143& $C_{2V}$ & -70.64347\\         
 94& $C_{1} $ & -43.10534  &119& $C_{S} $ & -56.78493  &144& $C_{3V}$ & -71.24204\\         
 95& $C_{1} $ & -43.63668  &120& $C_{1} $ & -57.25183  &145& $C_{2V}$ & -71.84058\\         
 96& $C_{1} $ & -44.15660  &121& $C_{1} $ & -57.81830  &146& $C_{5V}$ & -72.43938\\         
 97& $C_{1} $ & -44.72345  &122& $C_{1} $ & -58.41161  &147& $I_{H} $ & -73.03843\\         
 98& $T_{D} $ & -45.30545  &123& $C_{S} $ & -58.98351  &148& $C_{S} $ & -73.42275\\         
 99& $C_{2V}$ & -45.88888  &124& $C_{S} $ & -59.57674  &149& $C_{S} $ & -73.89112\\         
100& $C_{S} $ & -46.41998  &125& $C_{S} $ & -60.10860  &150& $C_{3V}$ & -74.44252\\                             
101& $C_{2V}$ & -46.95094  &126& $C_{1} $ & -60.61249  &   &          &          \\                        
102& $C_{2V}$ & -47.44697  &127& $C_{2V}$ & -61.20664  &   &          &          \\                        
103& $C_{S} $ & -47.98051  &128& $C_{1} $ & -61.77768  &   &          &          \\                        
                
\end{tabular}                         
\end{ruledtabular} 
\end{table*}

\end{document}